\documentclass{kapproc} % Computer Modern font calls

\usepackage{epsfig}

\normallatexbib

\begin{document}

%------------ article title  ------------------->>

\articletitle{Spectral density functional approach to electronic correlations and
magnetism in crystals}

%% optional, to supply a shorter version of the title for the running head:
\chaptitlerunninghead{SDF approach to electronic correlations}

\author{A. I. Lichtenstein}
\affil{NSRIM, University of Nijmegen\\
NL-6525 ED Nijmegen, The Netherlands}
\email{A.Lichtenstein@sci.kun.nl}

\author{M. I. Katsnelson}
\affil{Institute of Metal Physics\\
620219 Ekaterinburg , Russia}
\email{Mikhail.Katsnelson@usu.ru}

\author{G. Kotliar}
\affil{Serin Physics Laboratory, Rutgers University\\
Piscataway, New Jersey 08855, USA}
\email{kotliar@physics.rutgers.edu}

%% multiple authors may be separated with \\
%% \author{Samuel Bostaph\\
%% and Gregor Kariotis}

% optional prologue
%\prologue{<text>}{<author, year>}

% optional abstract
\begin{abstract}
A novel approach to electronic correlations and magnetism of crystals based
on realistic electronic structure calculations is reviewed. In its
simplest form it is a combination of the ``local density approximation'' (LDA)
and the dynamical mean field theory (DMFT) approaches. Using numerically
exact QMC solution to the effective DMFT multi-orbital quantum-impurity problem,
a successful description of electronic structure and finite temperature
magnetism of transition metals has been achieved. We discuss a simplified
perturbation LDA+DMFT scheme which combines the T-matrix and
fluctuation-exchange approximation (TM-FLEX). We end with a discussion of
cluster generalization of the non-local DMFT scheme and its applications to
the magnetism and superconductivity of high-T$_{c}$ superconductors.
\end{abstract}

% optional keywords
% \begin{keywords}
% Text, text...
% \end{keywords}

%------------ body of article ------------------->>

\section{Introduction}

The theory of electronic structure and magnetism of solids historically was
split into two distinct parts, namely, the model investigations of many-body
effects and the calculations of the energy spectra and properties of
specific compounds in the framework of density functional (DF) scheme \cite
{DF,DFrev}. Recently, within the dynamical mean-field theory (DMFT, for a
review see Ref. \cite{GKKR}) the correlation effects have been incorporated
into realistic electronic structure calculations [4-10]. 
%\cite{AnisDMFT,lda++,spflex,exchplus,chitra,voll,kotsav}.
This method has been
successfully applied to a number of classical problems of solid state
physics such as the finite-temperature magnetism of iron-group metals \cite
{FeNi}, $\alpha -\delta $ transition in plutonium \cite{Pu}, electronic
structure of doped Mott insulators \cite{voll}. In contrast with standard DF
theory, in this new approach known as ``LDA+DMFT'' \cite{AnisDMFT} or
``LDA++'' \cite{lda++} the total energy of the system (or, more accurately,
the thermodynamic potential $\Omega $) is considered as a functional of the
Green function instead of the denstity matrix \cite
{exchplus,chitra,berlin,SDF}. To stress this new feature more explicitely we
will use the term ``spectral density functional (SDF)''. The analytical
properties of the Green function garantee that the knowlegde of the spectral
density is equivalent to the knowledge of the time-dependent Green function
whereas the density matrix is just static value of the latter \cite{AGD}.
Here we will describe the basic ideas of the SDF method, both in the framework
of a standard DMFT and from a more general point of view, discuss possible
cluster generalizations of the DMFT, and consider the applications of SDF to
correlation effects and magnetism in transition metals.

\section{Dynamical Mean Field Theory: an effective action
perspective}

\bigskip A most economical approach to unifying the various dynamical mean-field
approximations in use, is provided by the effective action
construction \cite{fukuda}. The idea is to select a
set of variables which is relevant to the physics of the problem and to
write down a functional of the relevant variables. The extremum of this
functional yields the values of those variables in equilibrium, and the value
of this functional at stationarity gives the free energy of the system in
equilibrium.

Density functional theory is the simplest example of this construction
\cite{fernando}, here the total energy of the solid is expressed
in terms of the density of the electrons. Another well tested example is the
spin density functional theory in which the total energy is expressed in
terms of the spin and charge densities. The construction of an explicit
expression of the exact effective-action functional is usually not
available except for the case of very simple examples and the success of the
method relies on the availability of good approximations to this functional.
The LDA and the LSDA approaches have been extraordinarily successful for
weakly correlated systems.

The dynamical mean-field approach to model Hamiltonians of strongly
correlated systems on a lattice can also be brought to such an effective
action perspective.
More important this construction can be easily generalized to incorporate
the so-called extended dynamical mean-field approach (E-DMFT) \cite{edmft}, by
constructing an effective action for both the local Green function and the
local density-density and local spin-spin autocorrelation function.  The
effective-action approach also allows the formulation of the E-DMFT
approach in the case where spatial and spin or charge symmetries are broken.
\cite{chitra,pankov}.

The effective-action approach, allows us a simple combination of the density
functional theory and the DMFT. By Legendre transformation techniques one
can construct a functional of the density and of the local spectral function
of the heavy orbitals whose extremization would yield the exact density
local spectral function of the heavy orbitals, and the total energy of the
system. Again the exact form of this functional are not known in explicit
form, but useful approximations to it are available, and can be used to study
interesting problems.

This realistic DMFT, LDA+DMFT approach or LDA++ approach, was first
implemented ignoring the coupling between the density and the local spectral
function. This amounts to performing first a LDA calculation to derive a
tight-binding model Hamiltonian, and then performing a DMFT calculation for
the spectra of the DMFT Hamiltonian. This is very close in spirit to the
philosphy of model Hamiltonian calculations. Recently a full implementation
of the self-consistent determination of the density and the local spectra
was carried out by S. Savrasov \cite{Pu}, and this is now closer in spirit
to traditional first-principles electronic structure calculations.

The effective action approach, can be generalized to clusters, if short-range
 correlations need to be taken into account. In the context of model
Hamiltonians, a functional of the restriction of the Green function to a
given cluster is defined. The extremization of this functional give rise to
cluster dynamical mean field equations. There is no difficulty in
constructing hybrid funtionals for the self-consistent determination of both
the density and the cluster Green functions,  namely a CDMFT+LDA method.

In the next section we motivate the approximate DMFT form of those
functionals from a perspective of a reduced fermionic description.

\section{Fermionic reduced description and dynamical mean-field theory}

In the SDF approach two-electron subsytems are introduced, one of them is
described by the standard DF theory (usually this is the subsystem of $sp$%
-electrons) and dynamical interelectron correlations are taken into account
for the another one (usually this is the subsystem of $d$- or $f$%
-electrons). Further simplifications can be connected with the local form of
interelectron interactions (only Hubbard-type on-site electron correlations
are considered), using some approximations like DMFT or its cluster
generalizations \cite{jarrell,LK,cellular}, etc. From a general point of
view all these approaches can be considered as specific cases of a ``coarse
graining'' (reduced description)
ideology \cite{CG} when all the variables describing the system
can be separated into the ``gross'' and ``slave'' variables; the only
assumption that a closed set of the equations of motion for the gross
variables exist is sufficient to find an explicit form of these equations
\cite{zubarev}. Here we demonstrate the coarse graining procedure for
fermionic degrees of freedom for the system of interacting electrons in a
crystal.

Let us start with the functional integral over the Grassman creation and
annihilation electron fields $c^{+},c$, where the ``measure'' is $\exp
\left( -S\right)$,

\begin{equation}
S=-\int dx \int\limits_{0}^{\beta }d\tau \left[ c^{+}\left(x,\tau\right) \left(
\frac{\partial }{\partial \tau }+\mu -H_{0}\right) c\left(x, \tau \right) %
\right] +S_{int}\left[ c^{+},c\right]  \label{action}
\end{equation}
Here $\mu $ is the chemical potential, $\beta =1/T$ is the inverse
temperature, and $H_{0}$ is the ``one-particle'' part of the Hamiltonian. In the
SDF approach
\begin{equation}
H_{0}=-\nabla ^{2}/2+V_{KS}  \label{hamil1}
\end{equation}
where $V_{KS}$ is the Kohn-Sham self-consistent potential \cite{DF}.

In the spirit of a reduced description approach \cite{CG} we introduce ``gross''
variables $d=f\cdot c$ (more explicitly, $d_{\mu }=\sum\limits_{j}f_{\mu
j}c_{j}$ where $j$ are site indices; e.g., in a standard DMFT $f_{\mu
j}=\delta _{\mu 0}\delta _{j0}$). \ The effective action for the gross
variables $S_{eff}$ \ is defined by introducing a delta-functional $\delta
\left( d-f\cdot c\right) \delta \left( d^{+}-f^{\ast }\cdot c^{+}\right) $
\cite{FS} into the functional integral.
\newpage
\begin{eqnarray}
\exp \left( -S_{eff}\right) =\frac{1}{Z}\int DcDc^{+}\exp \left( -S\right)*\nonumber \\ 
\int D\lambda D\lambda ^{+}\exp \left[ i\lambda \left( d^{+}-f^{\ast }\cdot
c^{+}\right) +i\lambda ^{+}\left( d-f\cdot c\right) \right]
\end{eqnarray}

Unfortunately, this functional integral cannot be estimated exactly and
approximations are needed. Here we consider only the simplest approximation
to the derivation of the effective action, we replace S$_{int}\left[ c^{+},c%
\right] \rightarrow S_{int}^{CG}\left[ d^{+},d\right] $ but compensate for
the omission of the interaction terms
away from those in the gross variables, namely in the medium by adding a
medium self-energy in the standard DMFT

\bigskip
\begin{equation}
S\rightarrow \widetilde{S}=S+\sum_{i\geq 0}\int\limits_{0}^{\beta
}\int\limits_{0}^{\beta }d\tau d\tau ^{\prime }c^{+}\left( \tau, i\right)
\Sigma \left( \tau -\tau ^{\prime },i,i\right) c\left( \tau ^{\prime
},i\right)
\end{equation}

In other words, the DMFT approach treats exactly the interaction terms {\it %
only \ in }those part of the interaction Lagrangian which can be written in
terms of gross variables, while {\it the rest of the interaction terms
are handled in a gaussian  approximation} by replacing the interaction terms
by a medium self-energy. 

Now the effective measure for the gross variables is given by 
\begin{eqnarray}
\exp \left( -S_{eff}\right) =\frac{1}{Z}\int DcDc^{+}\exp \left( -\widetilde{%
S}_{CG}+d^{+}\Sigma d\right)*\nonumber \\
\int D\lambda D\lambda ^{+}\exp \left[ i\lambda
\left( d^{+}-f^{\ast }\cdot c^{+}\right) +i\lambda ^{+}\left( d-f\cdot
c\right) \right]  \label{eff1}
\end{eqnarray}
where $Z=\int DcDc^{+}\exp \left( -\widetilde{S}_{CG}\right) ,\widetilde{S}%
_{CG}=S_{med}+S_{int}^{CG}\left[ d^{+},d\right] .$

Passing to the Matsubara frequencies \cite{AGD} one has
\begin{eqnarray}
&&  \nonumber \\
S_{med} &=&-T\sum\limits_{\omega _{n}j}c^{+}\left( i\omega _{n}j\right)
\left[ i\omega _{n}+\mu -H_{0}-\Sigma \left( i\omega _{n}\right) \right]
c\left( i\omega _{n}j\right)  \label{matsubara}
\end{eqnarray}

\bigskip

In the standard DMFT scheme the correlations between $d(f)$-electrons on a
central site $j=0$  are treated exactly while the correlations in the
medium are treated in the Gaussian approximation. It is sometimes stated in
the literature that this assumption violates the translational invariance of
the system.  The effective-action approach, is meant to construct a
functional of the selected variables, so by construction it is not
possible to discuss within this framework the issue of translation
invariance.  The ``self-consistency condition'' does the best possible
job within a Gaussian approximation to restore the equivalence of the
central site and the medium.

We show below that this reasoning in a cluster setting leads to the
``cellular DMFT'' approach \cite{cellular}. In principle, the correlations
between the gross and slave variables (e.g., the long-range part of the
Coulomb interaction) also can be taken into account in a close analogy with
the classical spin models \cite{CG}; the corresponding modification of the
DMFT approach will be considered elsewhere. In accordance with the general
scheme of the non-equilibriun statistical operator method \cite{zubarev}, we
add an ``auxiliary field'' conjugated to the Green function of the gross
variables, namely, the term $d^{+}\bar{\Sigma}d$.
These fields will restore the translational
invariance for the fermionic Green function (see below, Eq.(\ref{equal})).

\bigskip Calculating the Gaussian integral over $c,c^{+}$ one has
\begin{eqnarray}
\exp \left( -S_{eff}\right) =\exp \left( -S_{int}^{CG}\left[ d^{+},d\right]
+d^{+}\bar{\Sigma} d\right)*\nonumber \\
\int D\lambda D\lambda ^{+}\exp \left[ i\lambda
d^{+}+i\lambda ^{+}d-\lambda ^{+}f\cdot G_{med}\cdot f^{\ast }\lambda \right]
\label{eff2}
\end{eqnarray}
where the medium Green function reads
\begin{equation}
G_{med}=\left[ i\omega _{n}+\mu -H_{0}-\Sigma \left( i\omega _{n}\right) %
\right] ^{-1}.  \label{medium}
\end{equation}
At last, calculating the Gaussian integral over $\lambda ,\lambda ^{+}$ we
obtain
\begin{equation}
S_{eff}\left[ d^{+},d\right] =S_{int}^{CG}\left[ d^{+},d\right]
-T\sum_{\omega _{n}}d^{+}\left( i\omega _{n}\right) {\cal G}_{0}^{-1}\left(
i\omega _{n}\right) d\left( i\omega _{n}\right)  \label{eff3}
\end{equation}
where
\begin{equation}
{\cal G}_{0}^{-1}\left( i\omega _{n}\right) =\left( f\cdot G_{med}\cdot
f^{\ast }\right) ^{-1}+\bar{\Sigma}  \label{bath}
\end{equation}
The self-consistency condition that determines $\Sigma $ requires the
equality of the local Green function computed from
the reduced description and from the medium agree, i.e.,
\begin{equation}
\left\langle Td(\tau )d^{+}(\tau ^{\prime })\right\rangle _{S}=\left\langle
Td(\tau )d^{+}(\tau ^{\prime })\right\rangle _{S_{eff}}  \label{equal}
\end{equation}
Further, $\left\langle Tc(\tau )c^{+}(\tau ^{\prime })\right\rangle
_{S}=G_{med}$ (because of the replacement $S_{int}\left[ c^{+},c\right]
\rightarrow S_{int}^{CG}\left[ d^{+},d\right] $) and
\begin{equation}
\left\langle Td(\tau )d^{+}(\tau ^{\prime })\right\rangle _{S}=f\cdot
G_{med}\cdot f^{\ast }  \label{equal1}
\end{equation}
On the other hand, $\left\langle Td(\tau )d^{+}(\tau ^{\prime
})\right\rangle _{S_{eff}}={\cal G}$ satisfies the Dyson equation
\begin{equation}
{\cal G}^{-1}={\cal G}_{0}^{-1}-\Sigma _{CG}=\left( f\cdot G_{med}\cdot
f^{\ast }\right) ^{-1}+\bar{\Sigma}-\Sigma _{CG}  \label{bath1}
\end{equation}
and we have $\bar{\Sigma}=\Sigma _{CG}$ where $\Sigma _{CG}$ is a ``coarse-grained''
self-energy. The most natural choise is $\Sigma _{CG}=f^{\ast }\cdot \Sigma
\cdot f.$

To obtain the main equation of the DMFT \cite{GKKR} we have to choise $f$ as
a projection operator on the central site. In this case Eq.(\ref{bath1})
will take the desired form
\begin{equation}
{\cal G}_{0}^{-1}\left( i\omega _{n}\right) =\left[ \sum\limits_{{\bf k}}%
\frac{1}{i\omega _{n}+\mu -t\left( {\bf k}\right) -\Sigma \left( i\omega
_{n}\right) }\right] ^{-1}+\Sigma \left( i\omega _{n}\right)  \label{DMFT}
\end{equation}
where the quasimomentum ${\bf k}$ runs the Brillouine zone, $t\left( {\bf k}%
\right) $ is the Fourier transform of the Hamiltonian $H_{0}$ projected into
the subspace of $d(f)$-electrons.

To consider posiible cluster generalizations one can choose let $f=PU^{+}P$
where $P$ is the projection operator on the cluster and $U$ is an unitary
transformation of the variables in this cluster. Thus our result reads:
\begin{equation}
{\cal G}_{0}^{-1}\left( i\omega _{n}\right) =\left[ \left( \frac{L_{c}}{L}%
\right) ^{d}\sum\limits_{{\bf k}_{c}}U^{+}\frac{1}{i\omega _{n}+\mu -t\left(
{\bf k}_{c}\right) -\Sigma }U\right] ^{-1}+U\Sigma U^{+}  \label{cluster}
\end{equation}
where ${\bf k}_{c}$ runs the ``new'' supercell Brillouin zone, $L$ and $%
L_{c} $ are the sizes of the crystal and cluster, correspondingly. In terms
of $\Sigma _{CG}=U\Sigma U^{+},$%
\begin{equation}
{\cal G}_{0}^{-1}\left( i\omega _{n}\right) =\left[ \left( \frac{L_{c}}{L}%
\right) ^{d}\sum\limits_{{\bf k}_{c}}\frac{1}{i\omega _{n}+\mu -Ut\left(
{\bf k}_{c}\right) U^{+}-\Sigma _{CG}}\right] ^{-1}+\Sigma _{CG}
\label{clus1}
\end{equation}
As it was shown in Ref. \cite{cellular} this equation for general basis set
can be very useful in the optimisation of interaction problem within the
cluster or cellular DMFT scheme. Another version of the cluster DMFT \cite
{LK} will be described below.

\section{Spectral density versus density functionals}

In a standard DF theory the thermodynamic potential for non-correlated
conduction-``c'' electrons $\Omega ^{c}$ is represented as a functional of
the electron density $\rho \left( {\bf r}\right) $ which is, generally
speaking, a matrix in spin indices. Formally it can be represented as a
thermodynamic potential of the Kohn-Sham quasiparticles \cite{DF}, $\Omega
_{sp}$, minus the contribution of the so called ``double counted'' terms, $%
\Omega _{dc}$:
\begin{eqnarray}
\Omega ^{c} &=&\Omega _{sp}^{c}-\Omega _{dc}^{c}  \nonumber \\
\Omega _{sp}^{c} &=&-Tr\log [i\omega +\nabla ^{2}/2-V_{KS}]  \\
\Omega _{dc}^{c} &=&\int V_{KS}({\bf r})\rho ({\bf r})d{\bf r-}\int V_{ext}(%
{\bf r})\rho ({\bf r})d{\bf r-}\frac{1}{2}\int \frac{\rho ({\bf r})\rho (%
{\bf r}^{\prime })}{|{\bf r}-{\bf r}^{\prime }|}d{\bf r}d{\bf r}^{\prime
}-E_{xc}[\rho ] \nonumber 
\label{DensFun}
\end{eqnarray}
where $Tr=Tr_{\omega iL\sigma },Tr_{\omega }$ is the sum over Matsubara
frequencies $Tr_{\omega }...=T\sum\limits_{\omega }e^{i\omega 0^{+}}...,$ $%
\omega =\pi T\left( 2n+1\right) ,$ $n=0,\pm 1,...,$ $T$ is the temperature,
and $iL\sigma $ are site numbers ($i$), orbital quantum numbers ($L={l,m}$)
and spin projections $\sigma $, respectively, $V_{ext}({\bf r})$ is the
external potential, $E_{xc}[\rho ]$ is the exchange-correlation energy, and
the Kohn-Sham effective potential is defined as
\begin{equation}
V_{KS}({\bf r})=V_{ext}({\bf r})+\int \frac{\rho ({\bf r}^{\prime })}{|{\bf r%
}-{\bf r}^{\prime }|}d{\bf r}^{\prime }+\frac{\delta E_{xc}[\rho ]}{\delta
\rho ({\bf r})}.  \label{KS}
\end{equation}
In contrast with the standard density functional theory, the SDF approach
deals with the real dynamical quasiparticles for correlated ``d-electrons''
defined via local Green functions rather than with Kohn-Sham
``quasiparticles'' which are, strictly speaking, only auxiliary states to
calculate the total energy. Therefore, instead of working with the
thermodynamic potential $\Omega $ as a {\it density} functional we have to
start from its general expression in terms of an exact Green function \cite
{luttinger}

\begin{eqnarray}
\Omega ^{d} &=&\Omega _{sp}^{d}-\Omega _{dc}^{d}  \nonumber \\
\Omega _{sp}^{d} &=&-Tr\left\{ \ln \left[ \Sigma -G_{0}^{-1}\right] \right\}
\nonumber \\
\Omega _{dc}^{d} &=&Tr\Sigma G-\Phi  \label{SDF}
\end{eqnarray}
where $G,G_{0}$ and $\Sigma $ are an exact Green function, its bare value
and self-energy, $\Phi $ is the Luttinger generating functional (sum of the
all connected skeleton diagrams without free legs), respectively. A
complete SDF thermodynamic potential is equal to $\Omega =\Omega ^{c}+\Omega
^{d}.$ We have to keep in mind also the Dyson equation
\begin{equation}
G^{-1}=G_{0}^{-1}-\Sigma  \label{DYSON}
\end{equation}
and the variational identity

\begin{equation}
\Sigma =\frac{\delta \Phi }{\delta G}.  \label{var}
\end{equation}
When neglecting the quasiparticle damping, $\Omega _{sp}$ will be nothing
but the thermodynamic potential of ''free'' fermions but with exact
quasiparticle energies. Formal analogies between Eqs.(\ref{DensFun}) and (\ref
{SDF}), (\ref{KS}) and (\ref{var}) are obvious: the self-energy plays the
role of the Kohn-Sham potential (without the external potential) and the
Green function plays the role of the density matrix. As an example of this
correspondence one can prove \cite{exchplus} in the framework of the SDF an
useful identity known as the ``local force theorem'' basically in the same
way as it has been done within DF theory \cite{MA,LKG}.

For both parts of the thermodynamic potential, $\Omega ^{c}$ and $\Omega
^{d},$ the local force theorem is based on extremum properties with respect
to the variation of the density matrix and Green function, respectively. 
Further we will consider the contribution $\Omega ^{d}$ (omitting the index $%
d$ for brevity). In principle, fermionic reduced description scheme allows us to
combined conduction ``c'' and correlated ``d'' part of the total SDF: $%
\Omega =\Omega ^{c}+\Omega ^{d}$ \cite{SDF}.

\section{Effective exchange interactions}

Let us discuss the problem of calculation of effective exchange interactions
($J_{ij}$) in correlated systems. In principle the $J_{ij}$ parameters are
not well defined for arbitrary magnetic systems, and the traditional way to
study spin excitations related to the calculation of non-local frequency-dependent
spin-susceptibility \cite{GKKR,Karlsson}. In this case the
auxiliary space-time dependent magnetic field is added to the the
Hamiltonian: ${\bf {\sigma h}}({\bf r},\tau )$ and the second derivative of
the free-energy with respect to the magnetic field gives the interacting
spin-susceptibility: $\chi ^{-1}=\chi _{0}^{-1}-\Gamma $, where $\chi _{0}$
is an empty-loop susceptibility and $\Gamma $ is the vertex correction \cite
{GKKR,Karlsson}. Here we consider a simple approximation of ``rigid spin
rotation'' of spectral density for a small angle:

\begin{equation}
\delta {\bf e}_{i}=\delta {\bf \varphi }_{i}\times {\bf e}_{i}  \label{rot1}
\end{equation}
where ${\bf e}_{i}$ is a general direction of constrained effective
spin-dependent potential on site $i$ and $\delta {\bf \varphi }_{i}$ is a
rotation vector. In this case it is useful to write explicitly the spinor
structure of the self-energy and Green functions:
\begin{eqnarray}
\Sigma _{i} &=&\Sigma _{i}^{c}+{\bf \Sigma }_{i}^{s}{\bf {\sigma }}
\label{spin} \\
G_{ij} &=&G_{ij}^{c}+{\bf G}_{ij}^{s}{\bf {\sigma }}  \nonumber
\end{eqnarray}
where $\Sigma _{i}^{\left( c,s\right) }=\frac{1}{2}\left( \Sigma
_{i}^{\uparrow }\pm \Sigma _{i}^{\downarrow }\right) $, ${\bf \Sigma }%
_{i}^{s}=\Sigma _{i}^{s}{\bf e}_{i},$ ${\bf {\sigma }}=(\sigma _{x},\sigma
_{y},\sigma _{z})$ are Pauli matrices, $G_{ij}^{c}=\frac{1}{2}Tr_{\sigma
}(G_{ij})$ and ${\bf G}_{ij}^{s}=\frac{1}{2}Tr_{\sigma }(G_{ij}{\bf {\sigma }%
})$. We suppose that the bare Green function $G^{0}$ does not depend on spin
direction and all the spin-dependent terms including the Hartree-Fock terms
are incorporated in the self-energy. In the rigid spin approximation we
assume that the unit vector ${\bf e}_{i}$ does not depend on the energy and
orbital indices and represents the direction of the average local magnetic
moment on the site $i$. Note, that the thermodynamic potential $\Omega $ should
be considered as a constrained SDF which depends on ${\bf e}_{i}$ as on
external parameters (cf. Ref. \cite{constr}). Then the variation of the
thermodynamic potential with respect to small spin-rotation can be written
as
\begin{equation}
\delta \Omega =\delta ^{\ast }\Omega _{sp}+\delta _{1}\Omega _{sp}-\delta
\Omega _{dc}  \label{var2}
\end{equation}
where $\delta ^{\ast }$ is the variation without taking into account the
change of the ''self-consistent potential'' (i.e. self-energy) and $\delta
_{1}$ is the variation due to this change of $\Sigma $. Taking into account
Eq. (\ref{var}) it can be easily shown (cf. Ref. \cite{luttinger}) that
\begin{equation}
\delta _{1}\Omega _{sp}=\delta \Omega _{dc}=TrG\delta \Sigma  \label{var3}
\end{equation}
and hence
\begin{equation}
\delta \Omega =\delta ^{\ast }\Omega _{sp}=-\delta ^{\ast }Tr\ln \left[
\Sigma -G_{0}^{-1}\right]  \label{var4}
\end{equation}
which is an analog of the ``local force theorem'' in the density functional
theory \cite{LKG}. \

In the case of rigid spin rotation the corresponding variation of the
thermodynamic potential can be written as
\begin{equation}
\delta \Omega ={\bf V}_{i}\delta {\bf \varphi }_{i}  \label{rot2}
\end{equation}
where the torque ${\bf V}_{i}$ is equal to
\begin{equation}
{\bf V}_{i}=2Tr_{\omega L}\left[ {\bf \Sigma }_{i}^{s}\times {\bf G}_{ii}^{s}%
\right]  \label{torque}
\end{equation}
Based on the expansion of this expression (\ref{torque}) in a sum of
pairwise contributions one can obtain \cite{exchplus} useful formula for the
effective magnetic interactions:
\begin{equation}
J_{ij}=-Tr_{\omega L}\left( \Sigma _{i}^{s}G_{ij}^{\uparrow }\Sigma
_{j}^{s}G_{ji}^{\downarrow }\right)  \label{Jij}
\end{equation}
and, correspondingly, for the stiffness tensor of a ferromagnet:
\begin{equation}
D_{\alpha \beta }=-\frac{2}{M}Tr_{\omega L}\sum\limits_{{\bf k}}\left(
\Sigma ^{s}\frac{\partial G^{\uparrow }\left( {\bf k}\right) }{\partial
k_{\alpha }}\Sigma ^{s}\frac{\partial G^{\downarrow }\left( {\bf k}\right) }{%
\partial k_{\beta }}\right)  \label{D}
\end{equation}
where $M$ is the magnetic moment per unit cell. These results generalize the
LSDA expressions of Ref. \cite{LKG} to the case of correlated systems. Note
that passing from Eq.(\ref{torque}) to Eq.(\ref{Jij}) is not accurate, since
the exchange parameters are connected with the second variations of the $%
\Omega $-potential and use of the local force theorem can not be justified.
Eq. (\ref{Jij}) corresponds to the ``empty loop'' approximation neglecting
the vertex corrections. At the same time, for the stiffness the latter are
absent and Eq.(\ref{D}) appears to be exact provided that the self-energy
and three-leg vertex are local (as in the DMFT) \cite{berlin}.

The fact that vertex corrections to the spin stiffness are absent within
DMFT, is also suggested by the analogy between
electric and spin transport developed in Ref. \cite{larkin}. The spin-wave
stiffness can be obtained from the zero-frequency
limit of the spin conductivity. Within DMFT , the charge current-current
correlation function does not require vertex corrections \cite{khurrana}
and can therefore be obained directly from the convolution of the one
electron spectra. These arguments are independent of the spin
structure, and can therefore be used \ for the spin-wave stiffness.

In order to elucidate the approximation behind the expression for the
exchange parameters (Eq. (\ref{Jij})), we consider the energy of a spiral
magnetic configuration with the rigid rotation of the spinor-electron
operators by the polar angles $\theta $ and $\varphi $:

\[
c_{im}\rightarrow U\left( {\theta _{i},\varphi _{i}}\right) c_{im}
\]
where

\begin{equation}
U\left( {\theta ,\varphi }\right) =\left(
\begin{array}{cc}
{\cos \theta /2} & {\sin \theta /2\exp \left( {-i\varphi }\right) } \\
{-\sin \theta /2\exp \left( {i\varphi }\right) } & {\cos \theta /2}
\end{array}
\right) \newline
\end{equation}

\noindent assuming that $\theta_{i}$ = const and $\varphi_{i}$ = ${\bf qR_{i}}$ where
${\bf R_{i}}$ is the site-lattice vector. Since we take into account only on-site
correlation effects the interaction term in the Hamiltonian is invariant
under that transformation, and the change of the Hamiltonian is

\begin{equation}
\begin{array}{l}
\delta H={\sum\limits_{ij}{Tr_{m\sigma }{\left[ {t_{ij}c_{i}^{+}\left( {%
U_{i}^{+}U_{j}-1}\right) c_{j}}\right] }}}=\delta _{1}H+\delta _{2}H \\
\delta _{1}H=\sin ^{2}{\frac{{\theta }}{{2}}}{\sum\limits_{k}{Tr_{m\sigma }}}%
{\left[ {\left( {t\left( {\bf k}{+}{\bf q}\right) -t\left( {\bf k}\right) }%
\right) c_{{\bf k}}^{+}c}_{{\bf k}}\right] } \\
\delta _{2}H={\frac{{1}}{{2}}}\sin \theta {\sum\limits_{ij}{Tr_{m}}}{\left[ {%
t_{ij}c_{i\downarrow }^{+}c_{j\uparrow }}\right] }\left( {\exp \left( {i{\bf %
qR}_{i}}\right) -\exp \left( {i{\bf qR}_{j}}\right) }\right)
\end{array}
\end{equation}

Consider further the case of small $\theta $, we can calculate change of the
total energy to lowest order in $\theta $ corresponds to the first order
in $\delta _{1}H$ and the second order in $\delta _{2}H$:

\begin{eqnarray*}
\delta E &=&{\frac{{\theta ^{2}}}{{4}}\{{\sum\limits_{k}{{\left[ {t\left(
{\bf k}{+}{\bf q}\right) -t\left( {\bf k}\right) }\right] }n_{k}-}}}iTr_{m}{%
\int {{\frac{{d^{4}k}}{{\left( {2\pi }\right) ^{4}}}}{\left[ {t\left( {\bf k}%
{+}{\bf q}\right) -t\left( {\bf k}\right) }\right] }}}\ast  \\
&&{{\gamma \left( k,q\right) G_{\downarrow }\left( k+q\right) {\left[ {%
t\left( {\bf k}{+}{\bf q}\right) -t\left( {\bf k}\right) }\right] }%
G_{\uparrow }\left( k\right) \}}},
\end{eqnarray*}
where $n_{k}=Tr_{m\sigma }{\left\langle {c_{k}^{+}c_{k}}\right\rangle }$,
$q$ is a four-vector with component ({\bf q,}0), and $\gamma $ is the
three-leg vertex. Our main approximation is to neglect of the vertex
corrections ($\gamma $ = 1). In this case the previous equation takes the
following form:

\begin{eqnarray}
\delta E &=&-{\frac{{\theta ^{2}}}{{4}}}Tr_{m}\{i{\int {{\frac{{d^{4}k}}{{%
\left( {2\pi }\right) ^{4}}}}{\left[ {t\left( {\bf k}{+}{\bf q}\right)
-t\left( {\bf k}\right) }\right] \ast }}} \label{deltaE}\\
&&{{G_{\downarrow }\left( {\bf k}{+}{\bf q}\right) {\left[ {G_{\downarrow
}^{-1}\left( k+q\right) -G_{\uparrow }^{-1}\left( k\right) +t\left( {\bf k}{+%
}{\bf q}\right) -t\left( {\bf k}\right) }\right] }G_{\uparrow }\left( {k}%
\right) {\}}}} \nonumber
\end{eqnarray}

Using the following consequence of the Dyson equation:

\begin{equation}
t\left( {\bf k}{+}{\bf q}\right) -t\left( {\bf k}\right) =G_{\uparrow
}^{-1}\left( {k}\right) -G_{\downarrow }^{-1}\left( {k+q}\right) +\Sigma
_{\uparrow }\left( {E}\right) -\Sigma _{\downarrow }\left( {E}\right)
\end{equation}

\noindent one can rewrite Eq. (\ref{deltaE}) in the form: $\delta E={%
\frac{{\theta ^{2}}}{{4}}}{\left[ {J\left( {0}\right) -J\left( {q}\right) }%
\right] }$ with the exchange integrals corresponding to Eq. (\ref{Jij}). We
conclude that the expression for $J_{ij}$ is accurate if the vertex
corrections can be neglected. Note that in the limit of small {\bf q} this can
be justified rigorously, provided that the self-energy and three-leg scalar
vertex are local. Therefore, the expression for the stiffness constant of
the ferromagnet (Eq. (\ref{D})) appears to be exact in the framework of DMFT
\cite{berlin}.

It should be stressed that the exchange integrals discussed here are just
static characteristics connected with energy of inhomogeneous spin
configurations. They determine the frequencies of spin excitation in
itinerant-electron systems only under adiabatic (rigid-spin) approximation.
For more general consideration one should calculate ${\bf q}$- and $\omega$-
dependent spin susceptibility.

\section{Electron correlations and finite-temperature magnetism in
transition metals}

Now we describe the applications of the SDF approach to a classical problem
of finite-temperature magnetism of the iron-group transition metals. Despite
a lot of attempts starting from seminal works by Heisenberg and Frenkel (for
review of early theories see e.g. \cite{herring,vons,moriya}) we have yet no
complete quantitative theory describing their magnetic and spectral
properties. The reason is that to describe the properties of Fe, Co, and Ni
one has to solve the problem of taking into account moderately strong
electronic correlations where approaches developed both for weakly
correlated systems such as normal-group metals and to higly correlated
systems such as Mott insulators are not, generally speaking, reliable.There
were many attempts to introduce by some way correlation effects in band
structure calculations of these metals \cite
{liebsch,treglia,manghi,nolting,steiner}, not referring to numerous purely
model works. But the question of applicability of specific approximations
such as lowest order perturbation theory \cite{treglia,steiner}, moment
method \cite{nolting} or three-body Faddeev equations \cite{manghi} is not
clear and one needs a reliable approach which would be checked carefully for
model systems and demonstrate its applicability for moderately correlated
systems. It has been demonstrated in Ref. \cite{FeNi} that the {\it ab
initio }dynamical mean-field theory does give a very successful description
of both correlation effects in the electron energy spectra and the
finite-temperature magnetic properties of Fe and, especially, Ni. Here we
present the corresponding results.

We start with the LDA Hamiltonian in the tight-binding orthogonal LMTO
representation $H_{mm^{\prime }}^{LDA}({\bf k})$ \cite{OKA}, where $m$
describes the orbital basis set containing $3d$-, $4s$- and $4p$- states, and $%
{\bf k}$ runs over the Brillouin zone (BZ). The interactions are
parameterized by a matrix of screened local Coulomb interactions, $%
U_{mm^{\prime }}$, and a matrix of exchange constants, $J_{mm^{\prime }}$,
which are expressed in terms of two screened Hubbard parameters, $U$ and $J$,
describing the average Coulomb repulsion and the interatomic ferromagnetic
exchange, respectively. We use the values $U=2.3$ $(3.0)$ eV for Fe (Ni) and
the same value of the interatomic exchange, $J=0.9$ eV for both Fe and Ni, a
result of constrained LDA calculations \cite{Coulomb,AnisDMFT,lda++}. These
parameters, which are consistent with those of many studies result in a very
good description of the physical properties of Fe and Ni.

As was discussed above, the DMFT maps the many-body system onto a
multi-orbital quantum impurity, i.e. a set of local degrees of freedom in a
bath described by the Weiss field function ${\cal G}$. The impurity action
(here $n_{m\sigma }={c}_{m\sigma }^{+}c_{m\sigma }$ and ${\bf c}(\tau
)=[c_{m\sigma }(\tau )]$ is a vector of Grassman variables) is given by:
\begin{eqnarray}
&&S_{eff}=-\int_{0}^{\beta }d\tau \int_{0}^{\beta }d\tau ^{\prime }Tr[{\bf c}%
^{+}(\tau ){{\bf {\cal G}}}^{-1}(\tau ,\tau ^{\prime }){\bf c}(\tau ^{\prime
})]+  \nonumber \\
&&\frac{1}{2}\sum_{m,m^{\prime },\sigma }\int_{0}^{\beta }d\tau \lbrack
U_{mm^{\prime }}n_{\sigma }^{m}n_{-\sigma }^{m^{\prime }}+(U_{mm^{\prime
}}-J_{mm^{\prime }})n_{\sigma }^{m}n_{\sigma }^{m^{\prime }}]  \label{path}
\end{eqnarray}
It describes the spin, orbital, energy and temperature dependent
interactions of a particular magnetic $3d$-atom with the rest of the crystal
and is used to compute the local Greens function matrix:
\begin{equation}
{\bf G}_{\sigma }(\tau -\tau ^{\prime })=-\frac{1}{Z}\int D[{\bf c},{\bf c}%
^{+}]e^{-S_{eff}}{\bf c}(\tau ){\bf c}^{+}(\tau ^{\prime })  \label{pathint}
\end{equation}
($Z$ is the partition function) and the impurity self-energy ${{\bf {\cal G}}%
}_{\sigma }^{-1}(\omega _{n})-{\bf G}_{\sigma }^{-1}(\omega _{n})={\bf %
\Sigma }_{\sigma }(\omega _{n})$ .

The Weiss field function is required to obey the self-consistency condition (%
\ref{DMFT}), which can be specified for a given case as
\begin{equation}
{\bf G}_{\sigma }(\omega _{n})=\sum_{{\bf k}}[(i\omega _{n}+\mu ){\bf 1}-%
{\bf H}^{LDA}({\bf k})-{\bf \Sigma }_{\sigma }^{dc}(\omega _{n})]^{-1}
\label{BZI}
\end{equation}
The local matrix ${\bf \Sigma }_{\sigma }^{dc}$ is the sum of two terms, the
impurity self energy and a so-called ``double counting '' correction, $%
E_{dc} $ which is meant to subtract the average electron-electron
interactions already included in the LDA Hamiltonian. For metallic systems
we proposed the general form of dc-correction: ${\bf \Sigma }_{\sigma
}^{dc}\left( \omega \right) ={\bf \Sigma }_{\sigma }\left( \omega \right) -%
\frac{1}{2}Tr_{\sigma }{\bf \Sigma }_{\sigma }\left( 0\right) $. This is
motivated by the fact that the static part of the correlation effects are
already well described in the density functional theory. Only the $d$-part
of the self-energy is presented in our calculations, therefore ${\bf \Sigma }%
_{\sigma }^{dc}=0$ for $s$- and $p$- states as well as for non-diagonal $d-s$%
, $p$ contributions. In order to describe the finite temperature
ferromagnetism of transition metals we use the {\em non} spin-polarized LDA
Hamiltonian ${\bf H}^{LDA}({\bf k})$ and accumulate{\em \ all}
temperature-dependent spin-splittings in the self-energy matrix ${\bf \Sigma
}_{\sigma }^{dc}\left( \omega _{n}\right) $.

We used the impurity QMC scheme (see Appendix)\ for the solution of the
multiband DMFT equations \cite{Rozenberg}. The Hirsch discrete
Hubbard-Stratonovich transformations \cite{HirschHS,Hirsch} introduces $(2M-1)M$
auxiliary Ising fields $S_{mm^{\prime }}^{\sigma \sigma ^{\prime }}$ where $%
M $ is the orbital degeneracy of the $d$-states and calculate ${\bf G}_{\sigma
}(\tau )$ by an exact integration of the fermion degrees of freedom in the
functional integral (Eq.(\ref{pathint})) \cite{GKKR}. In order to sample
efficiently all the spin configurations in the multi-band QMC scheme, it is
important to use ``global'' spin-flips: $[S_{mm^{\prime }}^{\sigma \sigma
^{\prime }}]\rightarrow \lbrack -S_{mm^{\prime }}^{-\sigma -\sigma ^{\prime
}}]$ in addition to the local moves of the auxiliary fields. The number of
QMC sweeps was of the order of 10$^{5}$. A parallel version of the DMFT
program was used to sample the 45 Ising fields for $3d$-orbitals. We used
256 ${\bf k}$-points in the irreducible part of the BZ for the k
integration. 10 to 20 DMFT iterations were sufficient to achieve convergence
far from the Curie point. Due to the cubic symmetry of the bcc-Fe and fcc-Ni
lattices the local Green function is diagonal in the basis of real spherical
harmonics. The spectral functions for real frequencies were obtained from
the QMC data by applying the maximum entropy method \cite{MEM}.

\begin{figure}[tbp]
\centerline{\epsfig{file=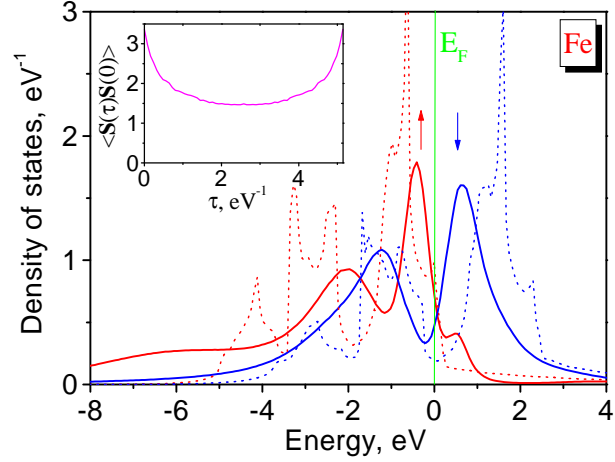, width=9cm,height=7cm,angle=0} }
\vskip 0.5cm
\caption{LDA+DMFT results for ferromagnetic iron ($T=0.8$ $T_{C}$).
The partial densities of d-states (full lines) is
compared with the corresponding LSDA results at zero temperature (dashed
lines) for the spin-up (arrow-up) and
spin-down (arrow-down) states.
The insert shows the spin-spin autocorrelation
function for T=1.2 $T_{C}$. }
\label{DOSFe}
\end{figure}

\begin{figure}[tbp]
\centerline{\epsfig{file=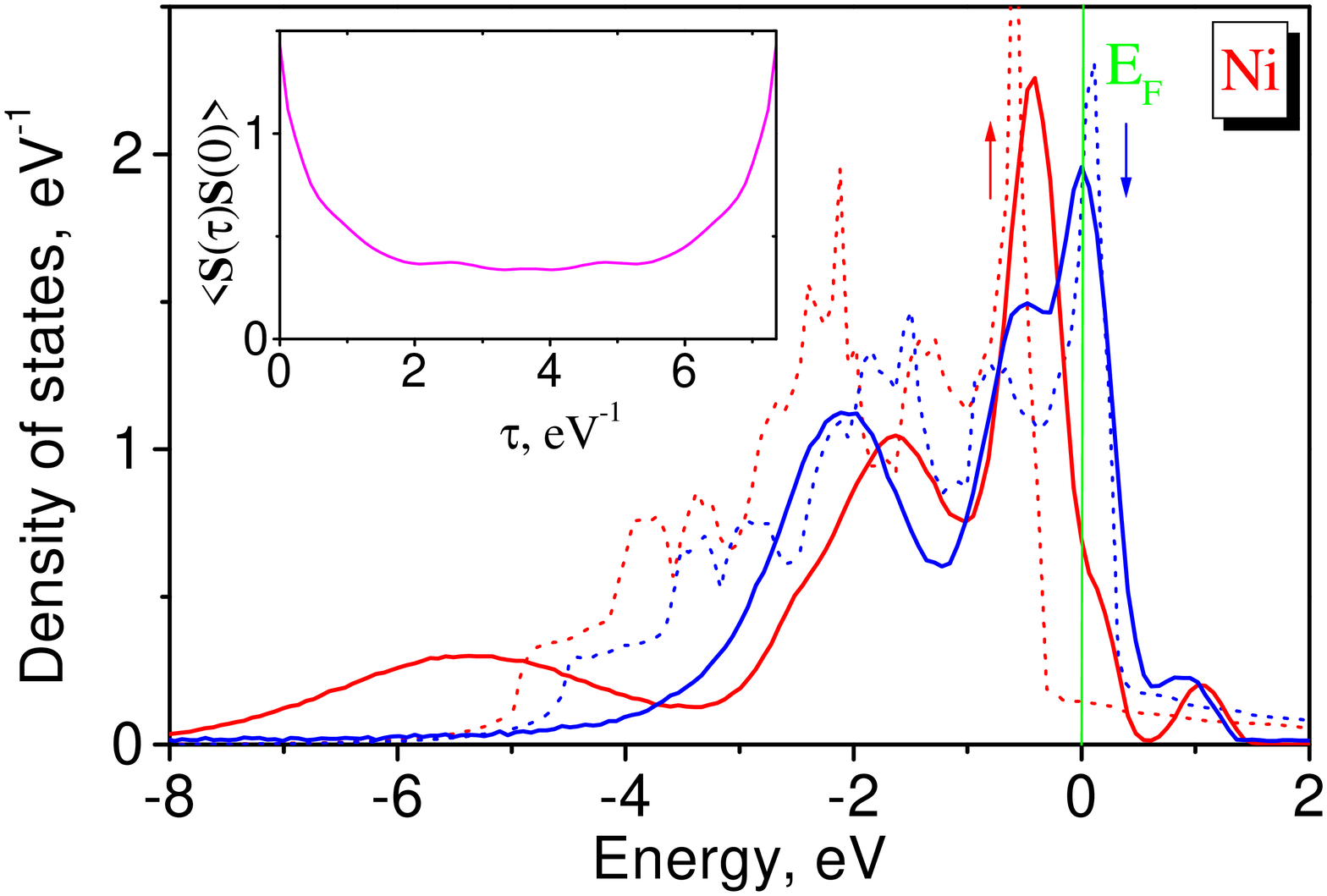, width=9cm,height=7cm,angle=0} }
\vskip 0.9cm
\caption{ Same quantities as in  Fig.1 for ferromagnetic nickel
($T=0.9$ $T_{C}$).
The insert shows the spin-spin autocorrelation
function for T=1.8 $T_{C}$. }
\label{DOSNi}
\end{figure}

Our results for the local spectral function for iron and nickel are shown in
Figs.\ref{DOSFe} and \ref{DOSNi},
respectively. The SDF approach describes well all the
qualitative features of the density of states (DOS), which is especially
non-trivial for nickel. Our QMC results reproduce well the three main
correlation effects on the one particle spectra below $T_{C}$ \cite
{satellite,satellite1,himpsel}: the presence of a famous 6 eV satellite, the
30\% narrowing of the occupied part of $d$-band and the 50\% decrease of
exchange splittings compared to the LDA results. Note that the satellite in
Ni has substantially more spin-up contributions in agreement with
photoemission spectra \cite{himpsel}. The exchange splitting of the d-band
depends very weakly on temperature from T=0.6 T$_{C}$ to T=0.9 T$_{C}$.
Correlation effects in Fe are less pronounced than in Ni, due to its large
spin-splitting and the characteristic bcc-structural dip in the density of
states for spin-down states near Fermi level, which reduces the density of
states for particle hole excitations.

Now we discuss the applications of the SDF approach to the description of
the finite-temperature magnetic properties of iron and nickel. While density
functional theory can in principle provide a rigorous description of the
thermodynamic properties, at present there is no accurate practical
implemention available. As a result the finite-temperature properties of
magnetic materials are estimated following a simple suggestion \cite{LKG},
whereby constrained DFT at $T=0$ is used to extract exchange constants for a
{\it classical} Heisenberg model, which in turn is solved using
approximation methods (e.g. RPA, mean field ) from classical statistical
mechanics of spin systems \cite{LKG,RJ,Eschrig,Antropov}. The most recent
implementation of this approach gives good values for the transition
temperature of iron but not of nickel \cite{kudrnovski}. While these
localized spin models give, by construction, at high temperatures a
Curie-Weiss like magnetic susceptibility, as observed experimentally in Fe
and Ni, they encounter difficulties in predicting the correct values of
the Curie constants\cite{gyorffy}.

\begin{figure}[tbp]
\centerline{\epsfig{file=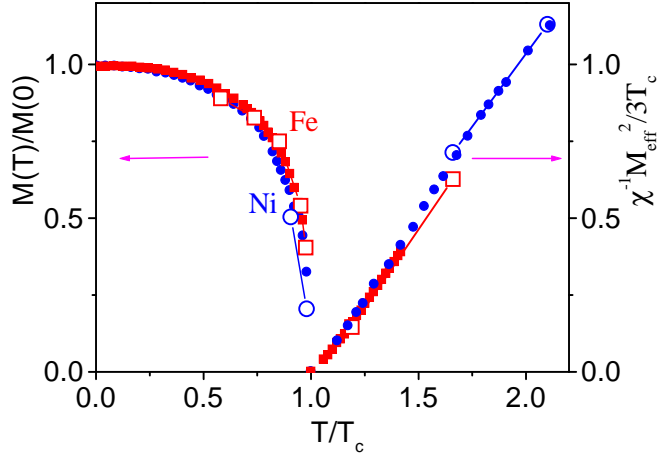, width=9cm,height=7cm,angle=0} }
\caption{Temperature  dependence of ordered  moment and the
inverse ferromagnetic susceptibility for Fe (open square) and Ni (open
circle) compared  with experimental results for Fe (square) and Ni
(circle) (from Ref.[33]). The calculated moments were normalized to the LDA
ground state magnetization (2.2 $\mu_B$ for Fe and 0.6 $\mu_B$ for Ni). }
\label{Mchi}
\end{figure}

The uniform spin susceptibility in the paramagnetic state, ${\chi _{q=0}}%
=dM/dH$, was extracted from the QMC simulations by measuring the induced
magnetic moment in a small external magnetic field. It includes the
polarization of the impurity Weiss field by the external field \cite{GKKR}.
The dynamical mean field results account for the Curie-Weiss law which is
observed experimentally in Fe and Ni. As the temperature increases above $%
T_{C}$, the atomic character of the system is partially restored resulting
in an atomic like susceptibility with an effective moment:
\begin{equation}
\chi _{q=0}=\frac{\mu _{eff}^{2}}{3(T-T_{C})}  \label{effective}
\end{equation}
The temperature dependence of the ordered magnetic moment below the Curie
temperature and the inverse of the uniform susceptibility above the Curie
point are plotted in Fig. \ref{Mchi} together with the corresponding experimental
data for iron and nickel\cite{wolfarth}. The LDA+DMFT calculations describes
the magnetization curve and the slope of the high-temperature Curie-Weiss
susceptibility remarkably well. The calculated values of high-temperature
magnetic moments extracted from the uniform spin susceptibility are $\mu
_{eff}=3.09$ $(1.50)\mu _{B}$ for Fe (Ni), in good agreement with the
experimental data $\mu _{eff}=3.13$ $(1.62)\mu _{B}$ for Fe (Ni)\cite
{wolfarth}.

%%%%% Fig.1 %%%%%%%%%%%%%%%%%%%%%%%%%%%%%%%%%%%%%
\vskip 0.5cm
%%%%%%%%%%%%%%%%%%%%%%%%%%%%%%%%%%%%%%%%%

%%%%% Fig.2 %%%%%%%%%%%%%%%%%%%%%%%%%%%%%%%%%%%%%
\vskip 0.9cm
%%%%%%%%%%%%%%%%%%%%%%%%%%%%%%%%%%%%%%%%%

We have estimated the values of the Curie temperatures of Fe and Ni from the
disappearance of spin polarization in the self-consistent solution of DMFT
problem and from the Curie-Weiss law in Eq.(\ref{effective}). Our
estimates $T_{C}=1900$ $(700)K$ are in reasonable agreement with
experimental values of $1043$ $(631)K$ for Fe (Ni) respectively\cite
{wolfarth}, considering the single-site nature of the DMFT approach, which
is not able to capture the reduction of $T_{C}$ due to long wavelength spin-waves. 
These effects are governed by the spin-wave stiffness. Since the
ratio of the spin-wave stiffness ($D$) to $T_{C}$, $T_{C}$/$a^{2}D$ is
nearly a factor of 3 larger for Fe than for Ni\cite{wolfarth} (a is the
lattice spacing), we expect the DMFT $T_{C}$ to be much higher than the
observed Curie temperature in Fe than in Ni. Note that this is a consequence
of the long-range oscilating character of exchange interactions in iron
compared to short-range ferromagnetic exchange interactions in nickel \cite
{kudrnovski}. Quantitative calculations demonstrating the sizeable reduction
of $T_{C}$ due to spin waves in Fe in the framework of a Heisenberg model
were performed in Ref \cite{kudrnovski}.
%%%%% Fig.3 %%%%%%%%%%%%%%%%%%%%%%%%%%%%%%%%%%%%%
\vskip 0.5cm
%%%%%%%%%%%%%%%%%%%%%%%%%%%%%%%%%%%%%%%%%
Within dynamical mean field theory one can also compute the local spin
susceptibility defined by
\begin{equation}
\chi _{loc}=\frac{g_{s}^{2}}{3}\int\limits_{0}^{\beta }d\tau \left\langle
{\bf S}\left( \tau \right) {\bf S}(0)\right\rangle  \label{local}
\end{equation}
where $g_{s}=2$ is the gyromagnetic ratio and ${\bf S}=\frac{1}{2}\sum_{m,\sigma
,\sigma ^{\prime }}c_{m\sigma }^{\dagger }{\bf \sigma }_{\sigma \sigma
^{\prime }}c_{m\sigma ^{\prime }}$ is single-site spin operator and ${\bf \
\sigma }=(\sigma _{x},\sigma _{y},\sigma _{z})$ are Pauli matrices. It
differs from the $q=0$ susceptibility by the absence of spin polarization
in the Weiss field of the impurity model. Eq.(\ref{local}) cannot be probed
directly in experiments but it is easily computed in DMFT-QMC. Its behavior
as function of temperature gives a very intuitive picture of the degree of
correlations in the system. In a weakly correlated system we expect Eq.(\ref
{local}) to be nearly temperature independent, while in a strongly
correlated system we expect a leading Curie-Weiss behavior at high
temperatures %\begin{equation}
$\chi _{local}={\mu _{loc}^{2}}/({3T+const})$ where $\mu _{loc}$ is an
effective local magnetic moment. In the Heisenberg model with spin $S$, $\mu
_{loc}^{2}=S(S+1)g_{s}^{2}$ and for well-defined local magnetic moments
(e.g., for rare earth magnets) this quantity should be temperature
independent. For the itinerant electron magnets, $\mu _{loc}$ is
temperature-dependent, due to a variety of competing many-body effects such
as Kondo screening, the induction of local magnetic moment by temperature
\cite{moriya} and thermal fluctuations which disorders the moments \cite{UFN}%
. All these effects are included in the DMFT calculations. The $\tau $%
-dependence of the correlation function $\left\langle {\bf S}\left( \tau
\right) {\bf S}(0)\right\rangle $ results in the temperature dependence of $%
\mu _{loc}$ and is displayed in the inserts on the Figs.\ref{DOSFe},\ref{DOSNi}.
Iron can be
considered as a magnet with very well-defined local moments above $T_{C}$
(the $\tau $-dependence of the correlation function is relatively weak),
whereas nickel is more of an itinerant electron magnet (stronger $\tau $
-dependence of the local spin-spin autocorrelation function).

The comparison of the values of the local and the $q=0$ susceptibility gives
a crude measure of the degree of short-range order which is present above $%
T_{C}$. As expected, the moments extracted from the local susceptibility Eq.(%
\ref{local}) are a bit smaller ( 2.8 \ $\mu _{B}$ for iron and 1.3 \ $\mu
_{B}$ for nickel) than those extracted from the uniform mangetic
susceptibility. This reflects the small degree of the short-range
correlations which remain well above $T_{C}$ \cite{SRO}. The
high-temperature LDA+DMFT clearly show the presence of a local-moment above $%
T_{C}$. This moment, is correlated with the presence of high energy features
(of the order of the Coulomb energies) in the photomeission. This is also
true below $T_{C}$, where the spin dependence of the spectra is more
pronounced for the satellite rigion in nickel than for that of the
quasiparticle bands near the Fermi level (Fig. \ref{DOSNi}). This can explain the
apparent discrepancies between different experimental determinations of the
high-temperature magnetic splittings \cite{LMM,sinkovic,aebi} as being the
results of probing different energy regions. The resonant photoemission
experiments \cite{sinkovic} reflect the presence of local-moment
polarization in the high-energy spectrum above the Curie temperature in nickel,
while the low-energy ARPES investigations \cite{aebi} results in
non-magnetic bands near the Fermi level. This is exactly the DMFT view on
the electronic structure of transition metals above $T_{C}$. Fluctuating
moments and atomic-like configurations are large at short times, which
results in correlation effects in the high-energy spectra such as
spin-multiplet splittings. The moment is reduced at longer time scales,
corresponding to a more band-like, less correlated electronic structure near
the Fermi level.

\section{Approximate solution for the self-energy: TM-FLEX method}

The QMC method described above is probably the most accurate way of solving
the effective impurity problem in the DMFT. However, it is rather cumbersome
and expensive computationally; besides that, it deals with the ``truncated''
two-indices interaction matrix (see Eq.(\ref{path})) instead of the complete
four-indices one. Therefore a scheme has been proposed in Ref. \cite{spflex}
based on a multiband spin-polarized generalization of the ``fluctuating
exchange'' (FLEX) approximation by Bickers and Scalapino \cite{flex}.
The original formulation of the FLEX approximation treats both
particle-hole (PH) and particle-particle (PP) channels on an equal footing. 
But their roles in
magnetism are completely different. The interaction of electrons with spin
fluctuations in PH channel leads to the most relevant correlation effects
\cite{moriya} whereas PP processes are important for the renormalizations of
the effective interactions in spirit of the $T$-matrix approach (``ladder
aprroximation'') by Galitskii \cite{tmatr} and Kanamori \cite{kanamori}.
Therefore we used in Ref. \cite{spflex} a ``two-step'' procedure when, at
first, the bare matrix vertex is replaced by a $T$-matrix, and, secondly, PH
channel processes with this effective interaction are taken into account
explicitly. This approximation has high enough accuracy both for the Hubbard
model and for real systems with moderate correlations $U <  W/2$ where 
$U$ is the Hubbard on-sire repulsion snergy and $W$ is the bandwidth (see
\cite{spflex} and Refs therein). However, specific form of the approximation
used in \cite{spflex} can be improved further by taking into account the
spin-dependence of the $T$-matrix. Here we present the formulation of this
T-matrix-FLEX approximation.

Let us start, as in Refs. \cite{lda++,spflex}, with the general many-body
Hamiltonian for a crystal in the LDA+U scheme \cite{revU}:

\begin{eqnarray}
H &=&H_{t}+H_{U}  \nonumber \\
H_{t} &=&\sum\limits_{\lambda \lambda ^{\prime }\sigma }t_{\lambda \lambda
^{\prime }}c_{\lambda \sigma }^{+}c_{\lambda ^{\prime }\sigma }  \nonumber \\
H_{U} &=&\frac{1}{2}\sum\limits_{\left\{ \lambda _{i}\right\} \sigma \sigma
^{\prime }}\left\langle \lambda _{1}\lambda _{2}\left| v\right| \lambda
_{1}^{\prime }\lambda _{2}^{\prime }\right\rangle c_{\lambda _{1}\sigma
}^{+}c_{\lambda _{2}\sigma ^{\prime }}^{+}c_{\lambda _{2}^{\prime }\sigma
^{\prime }}c_{\lambda _{1}^{\prime }\sigma \,,}  \label{hamilU}
\end{eqnarray}
where $\lambda =im$ are the site number $\left( i\right) $ and orbital $%
\left( m\right) $ quantum numbers, $\sigma =\uparrow ,\downarrow $ is the
spin projection, $c^{+},c$ are the Fermion creation and annihilation
operators, $H_{t}$ is the effective single-particle Hamiltonian from the
LDA, corrected for the double-counting of average interactions among
correlated electrons as it was described above, and the Coulomb matrix
elements are defined in the standard way

\begin{equation}
\left\langle 12\left| v\right| 34\right\rangle =\int d{\bf r}d{\bf r}%
^{\prime }\psi _{1}^{\ast }({\bf r})\psi _{2}^{\ast }({\bf r}^{\prime
})v\left( {\bf r-r}^{\prime }\right) \psi _{3}({\bf r})\psi _{4}({\bf r}%
^{\prime }),  \label{coulomb}
\end{equation}
where we define for briefness $\lambda _{1}\equiv 1$ etc. Following Ref. \cite
{tmatr} we take into account the ladder ($T$-matrix) renormalization of
the effective approximation:
\begin{eqnarray}
\left\langle 13\left| T^{\sigma \sigma ^{\prime }}\left( i\Omega \right)
\right| 24\right\rangle =\left\langle 13\left| v\right| 24\right\rangle -%
\frac{1}{\beta }\sum\limits_{\omega }\sum\limits_{5678}\left\langle 13\left|
v\right| 57\right\rangle *\nonumber \\   
G_{56}^{\sigma }\left( i\omega \right)
G_{78}^{\sigma ^{\prime }}\left( i\Omega -i\omega \right) \left\langle
68\left| T^{\sigma \sigma ^{\prime }}\left( i\Omega \right) \right| 24\right\rangle 
\label{tmatrix}
\end{eqnarray}
Further we rewrite the perturbation theory in terms of this effective
interaction matrix.

At first, we take into account the ``Hartree'' and ``Fock'' diagrams with
the replacement of the bare interaction by the $T$-matrix
\begin{eqnarray}
\Sigma _{12,\sigma }^{\left( TH\right) }\left( i\omega \right) &=&\frac{1}{%
\beta }\sum\limits_{\Omega }\sum\limits_{34\sigma ^{\prime }}\left\langle
13\left| T^{\sigma \sigma ^{\prime }}\left( i\Omega \right) \right|
24\right\rangle G_{43}^{\sigma ^{\prime }}\left( i\Omega -i\omega \right)
\nonumber \\
\Sigma _{12,\sigma }^{\left( TF\right) }\left( i\omega \right) &=&-\frac{1}{%
\beta }\sum\limits_{\Omega }\sum\limits_{34}\left\langle 14\left| T^{\sigma
\sigma }\left( i\Omega \right) \right| 32\right\rangle G_{34}^{\sigma
}\left( i\Omega -i\omega \right)  \label{HarFock}
\end{eqnarray}
Note that $\Sigma ^{(TH)}+$ $\Sigma ^{(TF)}$ contains exactly all the
second-order contributions. Now we have to consider the contribution of
particle-hole excitations to sigma. Similar to \cite{spflex} we will replace
in the corresponding diagrams the bare interaction by the static limit of the $T$%
-matrix. However, we improve the approximation \cite{spflex} by taking into
account its spin dependence. When considering the particle-hole channel we
replace in the Hamiltonian (\ref{hamilU}) $v\rightarrow T^{\sigma \sigma
^{\prime }}$ which is the solution of Eq.(\ref{tmatrix}) at $\Omega =0.$ Eq.
(\ref{HarFock}) is exact in the limit of low electron (or hole) density
which is important for the criterion of magnetism e.g. in the case of nickel
(with almost completely filled $d$- band).

Now we rewrite the effective Hamiltonian (\ref{hamilU}) with the replacement
$\left\langle 12\left| v\right| 34\right\rangle $ by $\left\langle 12\left|
T^{\sigma \sigma ^{\prime }}(\Omega=0)\right| 34\right\rangle $ in $H_{U}$.
To consider the correlation effects due to PH channel we have to separate
density ($d$) and magnetic ($m$) channels as in \cite{flex}

\[
d_{12}=\frac 1{\sqrt{2}}\left( c_{1\uparrow }^{+}c_{2\uparrow
}+c_{1\downarrow }^{+}c_{2\downarrow }\right)
\]

\begin{eqnarray}
m_{12}^{0} &=&\frac{1}{\sqrt{2}}\left( c_{1\uparrow }^{+}c_{2\uparrow
}-c_{1\downarrow }^{+}c_{2\downarrow }\right)  \nonumber \\
m_{12}^{+} &=&c_{1\uparrow }^{+}c_{2\downarrow }  \nonumber \\
m_{12}^{-} &=&c_{1\downarrow }^{+}c_{2\uparrow }\,,  \label{chan}
\end{eqnarray}
Then the interaction Hamiltonian can be rewritten in the following matrix
form
\begin{equation}
H_{U}=\frac{1}{2}Tr\left( D^{+}\ast V^{\parallel }\ast D+m^{+}\ast
V_{m}^{\perp }\ast m^{-}+m^{-}\ast V_{m}^{\perp }\ast m^{+}\right)
\label{hamnew}
\end{equation}
where * means the matrix multiplication with respect to the pairs of orbital
indices, e.g.
\begin{eqnarray*}
\left( V_{m}^{\perp }\ast m^{+}\right) _{11^{\prime }}
&=&\sum\limits_{34}\left( V_{m}^{\perp }\right) _{11^{\prime },22^{\prime
}}m_{22^{\prime }}^{+}, 
\end{eqnarray*}
the supervector D defined as  
\begin{eqnarray*}
D &=&\left( d,m^{0}\right) ,  D^{+}=\left(
\begin{array}{c}
d^{+} \\
m_{0}^{+}
\end{array}
\right) ,
\end{eqnarray*}
and the effective interactions have the following form: 
\begin{eqnarray}
\left( V_{m}^{\perp }\right) _{11^{\prime },22^{\prime }} &=&-\left\langle
12\left| T^{\uparrow \downarrow }\right| 2^{\prime }1^{\prime }\right\rangle
,  \nonumber \\
V^{\parallel } &=&\left(
\begin{array}{cc}
V^{dd} & V^{dm} \\
V^{md} & V^{dd}
\end{array}
\right) ,  \nonumber \\
V_{11^{\prime },22^{\prime }}^{dd} &=&\frac{1}{2}\sum\limits_{\sigma \sigma
^{\prime }}\left\langle 12\left| T^{\sigma \sigma ^{\prime }}\right|
1^{\prime }2^{\prime }\right\rangle -\frac{1}{2}\sum\limits_{\sigma
}\left\langle 12\left| T^{\sigma \sigma }\right| 2^{\prime }1^{\prime
}\right\rangle ,  \nonumber \\
V_{11^{\prime },22^{\prime }}^{mm} &=&\frac{1}{2}\sum\limits_{\sigma \sigma
^{\prime }}\sigma \sigma ^{\prime }\left\langle 12\left| T^{\sigma \sigma
^{\prime }}\right| 1^{\prime }2^{\prime }\right\rangle -\frac{1}{2}%
\sum\limits_{\sigma }\left\langle 12\left| T^{\sigma \sigma }\right|
2^{\prime }1^{\prime }\right\rangle ,   \\%
V_{11^{\prime },22^{\prime }}^{dm} &=&V_{22^{\prime },11^{\prime }}^{md}=\nonumber \\%
&&\frac{1}{2}\left[
\begin{array}{c}
\left\langle 12\left| T^{\uparrow \uparrow }\right| 1^{\prime }2^{\prime
}\right\rangle -\left\langle 12\left| T^{\downarrow \downarrow }\right|
1^{\prime }2^{\prime }\right\rangle -\left\langle 12\left| T^{\uparrow
\downarrow }\right| 1^{\prime }2^{\prime }\right\rangle + \nonumber \\ 
\left\langle 12\left| T^{\downarrow \uparrow }\right| 1^{\prime }2^{\prime }\right\rangle
-\left\langle 12\left| T^{\uparrow \uparrow }\right| 2^{\prime }1^{\prime
}\right\rangle +\left\langle 12\left| T^{\downarrow \downarrow }\right|
2^{\prime }1^{\prime }\right\rangle
\end{array}
\right]  \label{effpotent}
\end{eqnarray}
To calculate the PH contribution to the electron self-energy we first have
to write the expressions for the generalized susceptibilities, both
transverse $\chi ^{\perp }$ and longitudinal $\chi ^{\parallel }.$
The corresponding expressions are the same as in \cite{spflex} but with another
definition of the interaction vertices. One has

\begin{equation}
\chi ^{+-}(i\omega )=\left[ 1+V_m^{\perp }*\Gamma ^{\uparrow \downarrow
}(i\omega )\right] ^{-1}*\Gamma ^{\uparrow \downarrow }(i\omega )\,,
\label{xi+-}
\end{equation}
where

\begin{equation}
\Gamma _{12,34}^{\sigma \sigma ^{\prime }}\left( \tau \right)
=-G_{23}^\sigma \left( \tau \right) G_{41}^{\sigma ^{\prime }}\left( -\tau
\right)  \label{gamma}
\end{equation}
is an ``empty loop'' susceptibility
and $\Gamma (i\omega )$ is its Fourier transform. The
corresponding longitudinal susceptibility matrix has a more complicated form:

\begin{equation}
\chi ^{\parallel }(i\omega )=\left[ 1+V^{\parallel }*\chi _0^{\parallel
}(i\omega )\right] ^{-1}*\chi _0^{\parallel }(i\omega ),  \label{xipar}
\end{equation}
and the matrix of bare longitudinal susceptibility:

\begin{equation}
\chi _{0}^{\parallel }=\frac{1}{2}\left(
\begin{array}{cc}
\Gamma ^{\uparrow \uparrow }+\Gamma ^{\downarrow \downarrow }\, & \,\Gamma
^{\uparrow \uparrow }-\Gamma ^{\downarrow \downarrow } \\
\Gamma ^{\uparrow \uparrow }-\Gamma ^{\downarrow \downarrow }\, & \,\Gamma
^{\uparrow \uparrow }+\Gamma ^{\downarrow \downarrow }
\end{array}
\right) ,  \label{xi0par}
\end{equation}
in the $dd$-, $dm^{0}$-, $m^{0}d$-, and $m^{0}m^{0}$- channels ($d,m^{0}=1,2$
in the supermatrix indices). An important feature of these equations is the
coupling of longitudinal magnetic fluctuations and of density fluctuations.
It is absent in one-band Hubbard model due to the absense of the interaction
of electrons with parallel spins. For this case Eqs. (\ref{xi+-},\ref{xipar}%
) coinsides with the well-known result \cite{kubo}.

Now we can write the particle-hole contribution to the self-energy.
According to \cite{spflex} one has
\begin{equation}
\Sigma _{12,\sigma }^{(ph)}\left( \tau \right) =\sum\limits_{34,\sigma
^{\prime }}W_{13,42}^{\sigma \sigma ^{\prime }}\left( \tau \right)
G_{34}^{\sigma ^{\prime }}\left( \tau \right) ,  \label{sigph}
\end{equation}
with P-H fluctuation potential matrix:

\begin{equation}
W^{\sigma \sigma ^{\prime }}\left( i\omega \right) =\left[
\begin{array}{cc}
W^{\uparrow \uparrow }\left( i\omega \right) & W^{\perp }\left( i\omega
\right) \\
W^{\perp }\left( i\omega \right) & W^{\downarrow \downarrow }\left( i\omega
\right)
\end{array}
\right] ,  \label{wpp}
\end{equation}
where the spin-dependent effective potentials are defined as

\[
W^{\uparrow \uparrow }=\frac{1}{2}V^{\parallel }\ast \left[ \chi ^{\parallel
}-\chi _{0}^{\parallel }\right] \ast V^{\parallel }
\]

\[
W^{\downarrow \downarrow }=\frac{1}{2}V^{\parallel }\ast \left[ \widetilde{%
\chi }^{\parallel }-\widetilde{\chi }_{0}^{\parallel }\right] \ast
V^{\parallel }\
\]

\[
W^{\uparrow \downarrow }=V_{m}^{\perp }\ast \left[ \chi ^{+-}-\chi _{0}^{+-}%
\right] \ast V_{m}^{\perp }
\]

\[
W^{\downarrow \uparrow }=V_{m}^{\perp }\ast \left[ \chi ^{-+}-\chi _{0}^{-+}%
\right] \ast V_{m}^{\perp }.
\]
where $\widetilde{\chi }^{\parallel },\widetilde{\chi }_{0}^{\parallel }$
differ from $\chi ^{\parallel },\chi _{0}^{\parallel }$ by the replacement
of $\Gamma ^{\uparrow \uparrow }\Leftrightarrow \Gamma ^{\downarrow
\downarrow }$ in Eq.(\ref{xi0par}). We have subtracted the seconf-order
contributions since they have already been taken into account in Eq.(\ref{HarFock}%
).

Our complete expression for the self energy is
\begin{equation}
\Sigma =\Sigma ^{(TH)}+\Sigma ^{(TF)}+\Sigma ^{(PH)}  \label{sigmatot}
\end{equation}
This expression takes into account accurately spin-polaron effects because
of the interaction with magnetic fluctuations \cite{spflex}, the energy
dependence of $T$-matrix which is important for describing the satellite
effects in Ni \cite{liebsch}, contains the exact second-order terms in $v$ and
is rigorous (because of the first term) for almost filled or almost empty
bands. In spirit of the DMFT approach we have to use ${\cal G}_{0}$ instead
of $G$ in all the expressions when calculating the self-energy on a separated
central site. It should be noted that this TM-FLEX scheme is not conserved
(or ``$\Phi$-derivable'') therefore one need to inforce the Luttinger
theorem by introducing the $\mu_0$ for the bath Green function as in
iterative preturbation theory\cite{Kajueter}

We have started from the spin-polarized LSDA band structure of ferromagnetic
nickel within the TB-LMTO method \cite{OKA} in the minimal $s,p,d$ basis set
and used numerical orthogonalization to find the $H_{t}$ part of our
starting Hamiltonian. We take into accounts the Coulomb interactions only between
$d$-states. Semiempirical analysis of the appropriate interaction value
gives $U\simeq 2-4$ eV. The difficulties with choosing the correct value of $U$
are connected with complicated screening problems, definitions of orthogonal
orbitals in the crystal, and contributions of the intersite interactions. In
the quasiatomic (spherical) approximation the full $U$-matrix for the $d-$%
shell is determined by the three parameters $U,J$ and $\delta J$ or
equivalently by effective Slater integrals $F^{0}$, $F^{2}$ and $F^{4}$ \cite
{revU}. For example, $U= F^{0}$, $J=(F^{2}+F^{4})/14$ and we use the
simplest way of estimating $\delta J$ or $F^{4}$ keeping the ratio $F^{2}/%
F^{4}$ equal to its atomic value 0.625 \cite{revU}. Note that the value of
intra-atomic (Hund) exchange interaction $J$ is not sensitive to the
screening and approximately equals 0.9 eV in different estimations \cite
{revU}.

%%%% Fig.4 %%%%%%%%%%%%%%%%%%%%%%%%%%%%%%%%
\vskip 0.0cm
\begin{figure}[tbp]
\centerline{\epsfig{file=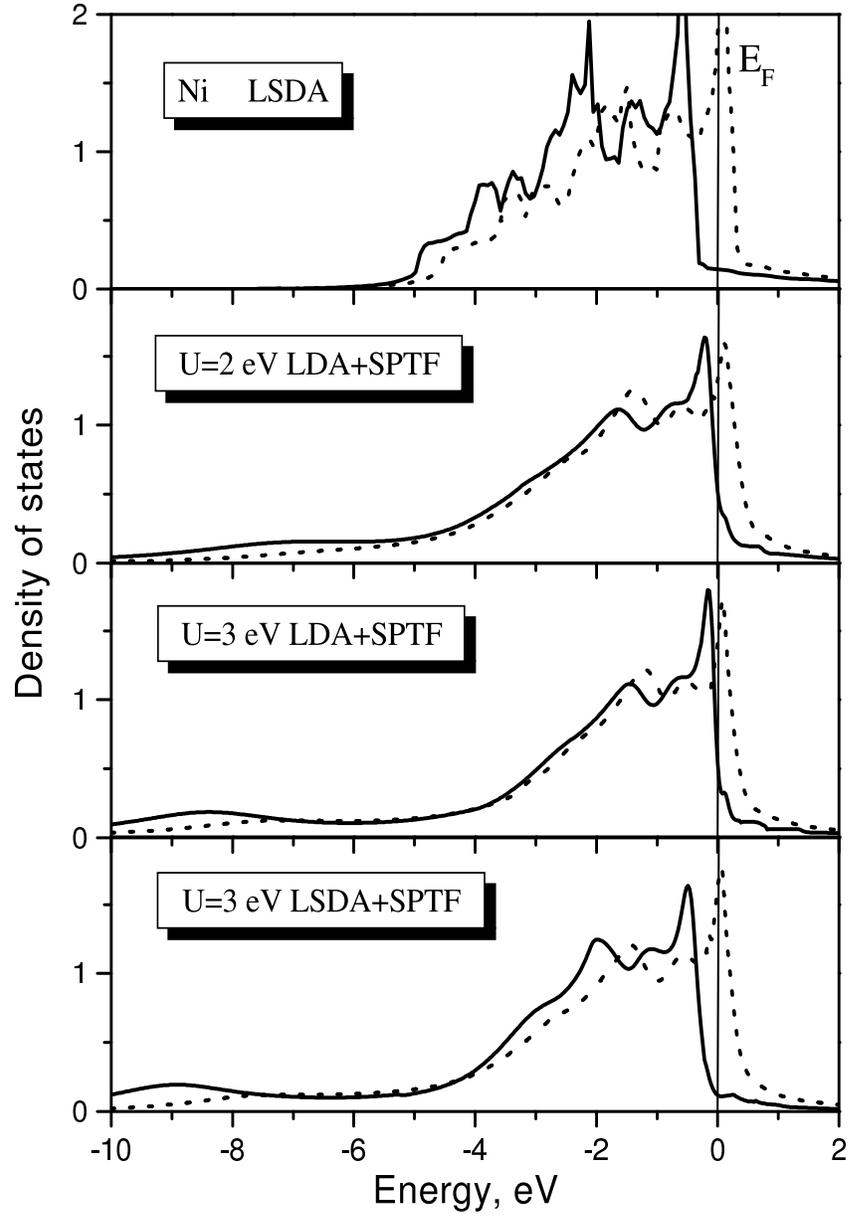, width=12cm,height=18cm,angle=0} }
\vskip 0.0cm
\caption{
Spin-up (full lines) and spin-down (dased lines) density of d-states
for ferromagnetic nickel in the LSDA
and the LDA+SPTF (LSDA+SPTF) calculations
for different average Coulomb interaction $U$ with $J=1$ eV
and temperature T=200~K.
}
\label{TMFLEX}
\end{figure}
%%%%%%%%%%%%%%%%%%%%%%%%%%%%%%%%%%%%%%%%%
The resulting DOS for Ni (Fig. \ref{TMFLEX}) shows that spin-polarized
TM-FLEX calculations approximatelly reproduce the satellite structure and
reduction of the band width in satisfactory agreement with exact
QMC-result(Fig. \ref{DOSNi})

\section{Cluster DMFT appoach: antiferromagnetism and superconductivity}

\label{periodic-cluster} As it was stressed above the SDF approach does not
necessarily connect with the standard DMFT scheme; one can use more
general choise of fermion coarse-grained variables, e.g., considering the
case of an effective cluster instead of an effective impurity. It is
especially important for the problems where intersite correlations are
involved from the beginning such as $d$-wave superconducting pairing \cite
{jarrell,LK} or charge ordering \cite{mazur}. Here we consider, following
Ref. \cite{LK}, one of the cluster generalizations of the DMFT, and its
application to the problem of magnetism and high-temperature
superconductivity (HTSC) of copper-oxide compounds.

The microscopic theory of high-temperature superconducting cuprates is still far
from a final understanding \cite{PWA,Scalapino,Pines}. One of the most
important recent experimental achievements was the discovery of the
pseudo-gap (PG) phenomenon above the superconducting transition temperature
\cite{PGAP} and existence of a sharp 41-meV resonance below T$_{c}$ related
to some collective antiferromagnetic excitations \cite{Neutr}. Thus, an
interplay of an antiferromagnetism (AFM) and $d$-wave superconductivity
(d-SC) in cuprates could be a natural way of discussing different HTSC
phenomena. This require a quantitative electronic structure theory including
two different type of the order parameters: AFM and d-SC. Within such
approach one can in principle analyze the phase diagram of HTSC compounds
and resolve the long-standing problem of competition between
antiferromagnetism and d-wave superconductivity in cuprates \cite
{SO5,Dagotto}.

\begin{figure}[t!]
\begin{center}
\vskip  0.5cm
\epsfig{figure=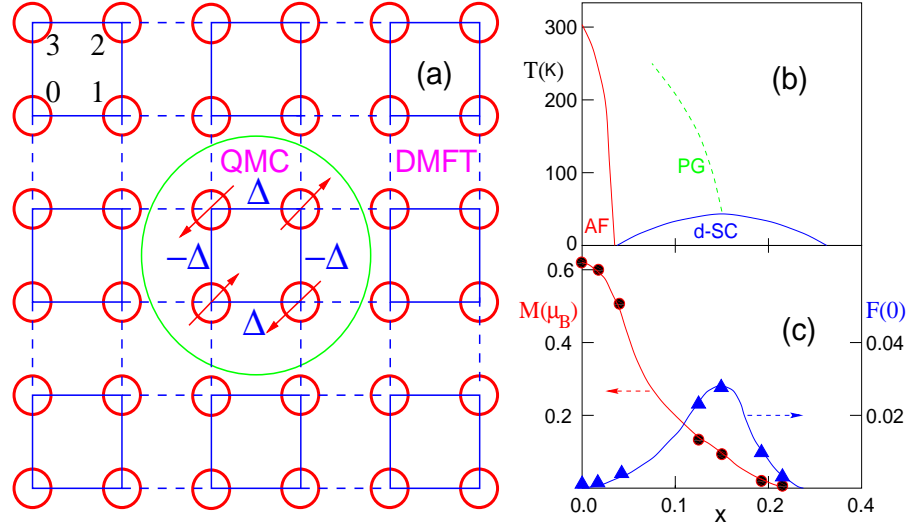,width=12cm,height=7cm,angle=0}
\vskip  0.4cm
\end{center}
\caption{
(a) Schematic representation of antiferromagnetic d-wave 2x2 periodically repeated clust
er;
(b) generic phase diagram of HTSC materials;
(c) The calculated values of two order parameters:
local magnetic moment M and d-SC equal time Green function $F^{01}(\tau=0)\equiv F(0)$
 for different  hole doping ($x$)
at the inverse temperature $\beta=60$ eV$^{-1} (T=190 K)$.
}
\label{AFMSC}
\end{figure}

A standard theoretical tool for cuprates electronic structure consists of
the two-dimensional Hubbard model \cite{PWA}. We start with the
extended-hopping Hubbard model on the square lattice:

\[
H=\sum_{ij}t_{ij}c_{i\sigma }^{+}c_{j\sigma }+\sum_{i}U_{i}n_{i\uparrow
}n_{i\downarrow }
\]
where $t_{ij}$ is an effective hopping and $U_{i}$ local Coulomb
interactions. We have chosen the nearest-neighbor hopping $t=0.25$ eV and
the next nearest hopping $t^{\prime }/t=-0.15$ for the model of La$_{2-x}$Sr$%
_{x}$CuO$_{4}$\cite{OKATB}. The total band width is $W$=2 eV and all Coulomb
parameters set to be $U$=1.2 eV ($U/W=0.6$). Let us introduce the
``super-site'' as an 2$\times $2 square plaquet. The numeration of the atoms
in the super-site is shown in the Fig.\ref{AFMSC}. It is useful to introduce
the superspinor $C_{i}^{+}=\{c_{i\alpha }^{+}\}$ where $\alpha =0,1,2,3$
(the spin indices are not shown). Taking into account the spin degrees of
freedom, this is the 8-component superspinor creation operator. Then the
crystal Green function for the Hubbard model can be rewritten as
\begin{equation}
G\left( {\bf k,}i\omega \right) =\left[ i\omega +\mu -h\left( {\bf k)-\Sigma
(}i\omega \right) \right] ^{-1}  \label{freeclust}
\end{equation}
where $h\left( {\bf k,}i\omega \right) $ is the effective hopping
supermatrix with self-energy corrections. For simplicity we will write
all the formulas in the nearest-neighbor approximations: \
\begin{equation}
h\left( {\bf k}\right) =\left(
\begin{array}{cccc}
0 & tK_{x}^{+} & 0 & tK_{y}^{+} \\
tK_{x}^{-} & 0 & tK_{y}^{+} & 0 \\
0 & tK_{x}^{-} & 0 & tK_{x}^{-} \\
tK_{y}^{-} & 0 & tK_{x}^{+} & 0
\end{array}
\right)  \label{hkom}
\end{equation}
where $K_{x(y)}^{\pm }=1+\exp \left( \pm ik_{x(y)}a\right) $, $a$ is the
lattice constant, and each element is a 2$\times $2 matrix in spin space.
Within the cluster-DMFT approach we introduce the intra-atomic self-energy $%
\Sigma _{0}$ and the inter-atomic self-energies $\Sigma _{x},$ $\Sigma _{y}$, and
both functions are of intra-site nature in the sense of our
super-site:
\begin{equation}
\Sigma \left( i\omega \right) =\left(
\begin{array}{cccc}
\Sigma _{0} & \Sigma _{x} & 0 & \Sigma _{y} \\
\Sigma _{x}^{\ast } & \Sigma _{0} & \Sigma _{y} & 0 \\
0 & \Sigma _{y}^{\ast } & \Sigma _{0} & \Sigma _{x}^{\ast } \\
\Sigma _{y}^{\ast } & 0 & \Sigma _{x} & \Sigma _{0}
\end{array}
\right)  \label{sigmamat}
\end{equation}

For the small 2$\times $2 cluster it is usefull to introduce the
translationally invariant ({\bf k}-dependent) self-energy and rewrite $%
h\left( {\bf k)+\Sigma (}i\omega \right) ->h\left( {\bf k,}i\omega \right) $
where
\begin{equation}
h\left( {\bf k,}i\omega \right) =\left(
\begin{array}{cccc}
\Sigma _{0} & t_{x}K_{x}^{+} & 0 & t_{y}K_{y}^{+} \\
t_{x}^{\ast }K_{x}^{-} & \Sigma _{0} & t_{y}K_{y}^{+} & 0 \\
0 & t_{y}^{\ast }K_{x}^{-} & \Sigma _{0} & t_{x}^{\ast }K_{x}^{-} \\
t_{y}^{\ast }K_{y}^{-} & 0 & t_{x}K_{x}^{+} & \Sigma _{0}
\end{array}
\right) .  \label{trinv}
\end{equation}

The effective Hamiltonian is defined through the renormalized energy
dependent hoppings: $t_{x}=$ $t+\Sigma _{x},$ $t_{y}=t+\Sigma _{y}$. The
functions $\Sigma _{0}\left( i\omega \right) ,$ $\Sigma _{x}\left( i\omega
\right) ,$ $\Sigma _{y}\left( i\omega \right) $ are found self-consistently
within the cluster DMFT scheme and for the $d$-wave superconduction state $%
\Sigma _{x}\neq \Sigma _{y}.$ It is straightforward to generalize this
scheme for a next-nearest neighbor hopping as well as the long-range Green
function and the self-energy. In this case we can renormalized also the
second-nearest hopping: $t_{xy}=t^{\prime }+\Sigma _{xy}$ for the 2$\times $%
2 cluster, where $\Sigma _{xy}$ (or $\Sigma ^{02}$) is the non-local
self-energy in the $xy$ direction.

%%%%Fig.5%%%%%%%%%%%%%%%%%%%%%%%%%%%
%%%%%%%%%%%%%%%%%%%%%%%%%%%%%%%%%%%%

In the cluster version of the DMFT scheme one can write the matrix equation
for the bath Green function matrix ${\cal G}$ which describe an effective
interaction with the rest of crystal:
\[
{\cal G}^{-1}\left( i\omega \right) =G^{-1}\left( i\omega \right) +\Sigma
\left( i\omega \right) ,
\]
where the local cluster Green function matrix is equal to $G_{\alpha \beta
}\left( i\omega \right) =$ $\sum\limits_{{\bf k}}G_{\alpha \beta }\left(
{\bf k,}i\omega \right) $ , and the summation is run over the Brillouin zone of
the square lattice. If instead of Eq.(\ref{trinv}) we use Eq.(\ref{hkom}),
it would corresponds to a free cluster DMFT scheme or the simplest case of
the so-called cellular-DMFT. Note that we use a translationally invariant
self-energy obtained from the cluster DMFT scheme (Eq.(\ref{trinv})) or the
simplest version of so-called cellular DMFT approach \cite{cellular}. We
believe that for a given choise of the small (2$\times $2) cluster, the
renormalization of the hopping parameter by $\Sigma _{x},$ $\Sigma _{y}$
terms is physically essential. The present ``matrix'' form of a cluster DMFT
with the self-energy which is not periodic inside the cluster allow us to
study a multicomponent ordered state. Unfortunately, in contrast with the
so-called Dynamical Cluster Approximation (DCA) \cite{jarrell} or \
cellular-DMFT approach \cite{cellular} (see above, Section 1), we were
unable to prove the casuality of this approach for arbitrary band structure and
interaction parameters. However, the casuality of the Green function for the
{\it realistic} choise of the parameters has been checked numerically.

In this case we have the standard DMFT problem with four ``orbital'' states
per super-site. It has been solved by the multi-orbital QMC technique
described above (Section 3). We used the generalized Nambu technique \cite
{Nambu} to analyze the coexistence of the magnetic ordering and
superconductivity. Let us introduce the superspinor
\[
\Psi _{i}^{+}(\tau )\equiv (\psi _{1i}^{+},\psi _{2i}^{+},\psi
_{3i}^{+},\psi _{4i}^{+})=\left( c_{i\uparrow }^{+},c_{i\downarrow
}^{+},c_{i\uparrow },c_{i\downarrow }\right)
\]
and the anomalous averages describing the (collinear) antiferromagnetism $%
\left\langle c_{i\uparrow }^{+}c_{j\downarrow }\right\rangle $ and the
superconductivity $\Delta _{ij}=$ $\left\langle c_{i\downarrow }c_{j\uparrow
}\right\rangle $.

The generalization of the Hirsch-Fye QMC-algorithm \cite{Hirsch} for
the superconducting problem \cite{Georges} have been used. In the 4-spinor case
a discrete Hubbard-Stratonovich transformation has the following form:

\begin{eqnarray}
\exp [-\Delta \tau U_{i}n_{i\uparrow }n_{i\downarrow }+\frac{\Delta \tau
U_{i}}{2}(n_{i\uparrow }+n_{i\downarrow })] =  \label{hirsuper} \nonumber \\
\frac{1}{2}\sum_{\sigma =\pm 1}\exp [\lambda _{i}\sigma (\psi
_{1i}^{+}\psi _{1i}-\psi _{2i}^{+}\psi _{2i}-\psi _{3i}^{+}\psi _{3i}+\psi
_{4i}^{+}\psi _{4i})]  
\end{eqnarray}
where $\lambda _{i}=\frac{1}{2}{\rm arccosh}[\exp (\frac{1}{2}\Delta \tau
U_{i})]$.

Since we take into account only the singlet pairing, we obtain the following
nonzero elements of the d-SC energy gap parameters: $\Delta =\Delta
_{01}=-\Delta _{12}=\Delta _{23}=-\Delta _{30}$. One can chose $\Delta _{ij}$
to be real and therefore symmetric: $\Delta _{ij}=\Delta _{ji}$. Separating
normal and anomalous parts of the Green function we have
\begin{equation}
G\left( {\bf k,}\tau ,\tau ^{\prime }\right) =\left(
\begin{array}{cc}
G\left( {\bf k,}\tau ,\tau ^{\prime }\right) & F\left( {\bf k,}\tau ,\tau
^{\prime }\right) \\
F^{+}\left( {\bf k,}\tau ,\tau ^{\prime }\right) & -G\left( -{\bf k,}\tau
^{\prime },\tau \right)
\end{array}
\right)  \label{Greenmatr}
\end{equation}
where $G\left( {\bf k,}\tau ,\tau ^{\prime }\right) =-\left\langle T_{\tau
}C_{{\bf k}}\left( \tau \right) C_{{\bf k}}^{+}\left( \tau ^{\prime }\right)
\right\rangle $ and  $F\left( {\bf k,}\tau ,\tau ^{\prime }\right)
=-\left\langle T_{\tau }C_{{\bf k}}\left( \tau \right) C_{-{\bf k}}\left(
\tau ^{\prime }\right) \right\rangle $ are the matrices in spin and
``orbital'' space. It is convenient to expand the anomalous Green function
in Pauli matrices $F=\left( F^{0}+{\bf F\sigma }\right) i\sigma ^{y}$ and
use the symmetry properties \cite{GV}:
\begin{eqnarray}
F^{0}\left( {\bf k,}\tau ,\tau ^{\prime }\right) &=&F^{0}\left( -{\bf k,}%
\tau ^{\prime },\tau \right)  \label{sym} \\
{\bf F\left( k,\tau ,\tau ^{\prime }\right) } &=&-{\bf F}\left( -{\bf k,}%
\tau ^{\prime },\tau \right) .  \nonumber
\end{eqnarray}
then a $4\times 4$ spinor formalism is reduced to $2\times 2$ one in the
collinear antiferromagnetic case with the d-wave superconductivity with the
following spin-matrix form of the local Green function for the super-site:

\begin{equation}
G\left( \tau ,\tau ^{\prime }\right) =\left(
\begin{array}{cc}
G_{\uparrow }\left( \tau ,\tau ^{\prime }\right) & F\left( \tau ,\tau
^{\prime }\right) \\
F(\tau ,\tau ^{\prime }) & -G_{\downarrow }\left( \tau ^{\prime },\tau
\right) ,
\end{array}
\right)  \label{GM1}
\end{equation}
and the QMC formalism for the antiferromagnetic superconducting state is
equivalent to the previous non-magnetic one \cite{Georges}. Using the
discretization of $[0,\beta ]$ interval with $L$-time slices: $\Delta \tau
=\beta /L$ ($\beta =1/T$ is an inverse temperature) the $G_{\sigma }$- and $%
F $- Green functions become the matrices of $2NL$ dimension, where $N$ is
the number of atoms in the cluster. After Fourier transform to the Matsubara
frequencies the Green function matrix has the following form:

\begin{equation}
G\left( i\omega \right) =\left(
\begin{array}{cc}
G_{\uparrow }\left( i\omega \right) & F\left( i\omega \right) \\
F(i\omega ) & -G_{\downarrow }^{\ast }\left( i\omega \right)
\end{array}
\right)  \label{GM2}
\end{equation}
In superconducting states the self-energy defined as \cite{GKKR}:

\begin{equation}
{\cal G}^{-1}\left( i\omega \right) -G^{-1}\left( i\omega \right) =\left(
\begin{array}{cc}
\Sigma _{\uparrow }\left( i\omega \right) & S\left( i\omega \right) \\
S(i\omega ) & -\Sigma _{\downarrow }^{\ast }\left( i\omega \right)
\end{array}
\right) ,  \label{GM3}
\end{equation}
and the inverse crystal Green function matrix is equal to:

\begin{equation}
G^{-1}\left( {\bf k,}i\omega \right) =\left(
\begin{array}{cc}
i\omega +\mu -h\left( {\bf k,}i\omega \right) & s\left( {\bf k,}i\omega
\right) \\
s\left( {\bf k,}i\omega \right) & i\omega -\mu +h^{\ast }\left( {\bf k,}%
i\omega \right)
\end{array}
\right)  \label{GM4}
\end{equation}
where $s\left( {\bf k,}i\omega \right) $ is the translationally invariant anomalous
part of the self-energy $S(i\omega )$ similar to Eq.(\ref{hkom}).

%%%%%%Fig.6 %%%%%%%%%%%%%%%%%%%%%%%%
\begin{figure}[t!]
\begin{center}
\vskip -0.5cm
\epsfig{figure=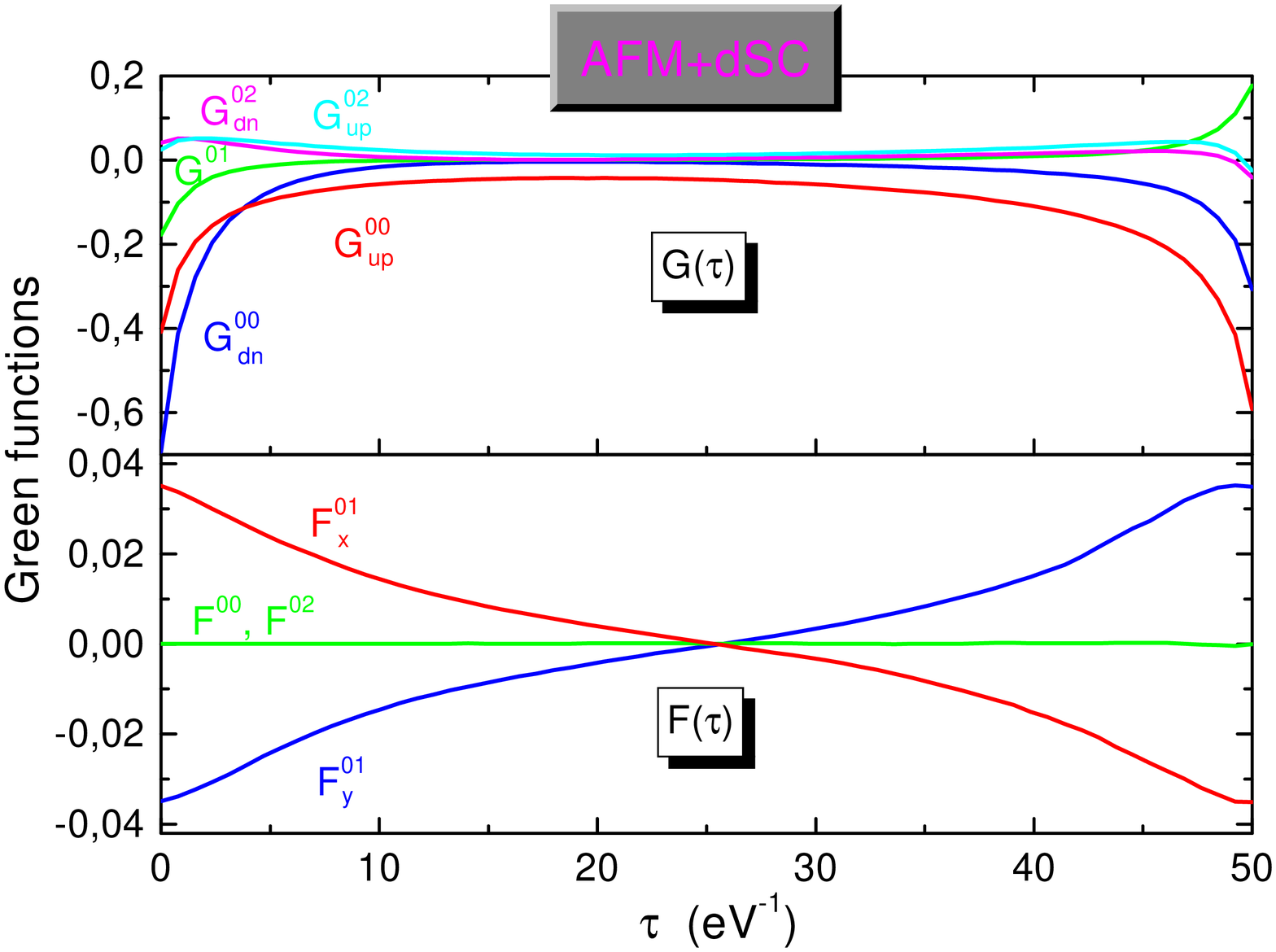,width=12cm,height=10cm,angle=0}
\vskip -0.4cm
\end{center}
\caption{
Imaginary time normal ($G_{\sigma}$) and superconducting ($F$) Green functions for the
2$\times$2 cluster DMFT solution  with second-nearest neighbor hopping
and inverse temperature $\beta=50$ eV$^{-1}$
(T=230 K).
}
\label{GTAU}
\end{figure}
%%%%%%%%%%%%%%%%%%%%%%%%%%%%%%%%%%%%

The two-component order parameters state which includes  Neel
antiferromagnetism and d-wave superconductivity (Fig.\ref{AFMSC}a) lowered the
symmetry of the effective cluster-DMFT problem. A self-consistent DMFT
cluster problem with AFM and d-SC general order parameters have been solved
within the QMC scheme for 8x8 matrix Green function with L=64 time slices.
The resulting two order parameters for $\beta =60$ eV$^{-1}$ (T=190 K) and $%
t^{\prime }=0$ presented in Fig.\ref{AFMSC}c together with the generic HTSC phase
diagram (Fig.\ref{AFMSC}b) as function of the hole doping. In this case the ordered
magnetic moment is directly related with imaginary-time Green function $%
G_{\sigma }(\tau )$: $M=G_{\uparrow }^{00}\left( 0\right) -G_{\downarrow
}^{00}\left( 0\right) $ and for the d-SC order parameter we chose a positive
value of superconducting imaginary time Green function $F^{01}\left(
0\right) $. It is important, that we find no serious sign-problem for all
QMC calculations with various doping level, probably due to ``stabilized''
antiferromagnetic dynamical mean fields acting on the atoms in our 2$\times$2
cluster. Note that the AFM cluster-DMFT solution exists for a much higher
doping concentration than experimental AFM ordered state and describes a
dynamical mean-field version of AFM-spin fluctuations related to pseudogap
phenomena (the PG-region on Fig.\ref{AFMSC}b). The maximum of d-SC order parameter
corresponds to a doping level of about 15$\%$ in agreement with the
generic HTSC phase diagram. The d-SC order parameter is zero close to the
undoped region ($x$=0), due to presence of a large AFM-gap. When the AFM-gap
is closed ($x\sim 5\%$) the d-SC states develope but for $x>20\%$ the d-gap
decrease again since AFM spin-fluctuations around ($\pi ,\pi $) point
disappear \cite{Scalapino}. The precise characteristic of the phase diagram
including the interactions between the AFM and d-SC order parameters demands an
extensive cluster-DMFT calculations for different temperatures and
doping.

We would like to note that the existing of d-SC cluster-DMFT solution for
such high temperatures does not necessary means that the superconducting
transition temperature is larger then 190K in our model. A crude estimation
shows that the d-SC solution disappears at $T$=300 K for $x$=0.15 and the AFM solutions
for x=0 become unstable at the temperature just above 1000 K. This could be
the sign of a ``local'' AFM solution and a local d-wave solution, like local
moments in magnetic systems \cite{spflex}. Due to the multiscale nature of the
problem under considerations, essentially different energies connected with
local moment formation, long-range magnetic order, local d-wave pairs within
the 2$\times$2 plaquet, and finally coherent superconductivity, it is difficult to
distinguish a real long-range ordering from slow dynamical fluctuations in
our QMC simulations. We plane to separate these energy scales analytically
and estimate superconductiong transition temperature in a future publication.

The role of next-nearest hopping is to lower of the van-Hove singularity
\cite{OKATB} which increases the density of state at the Fermi level for the
hole-doped case and favores the d-SC solution for a moderate correlation
strength. There is also a change in the spin-fluctuation spectrum related with
the broadening of AFM-peak near the ($\pi ,\pi $) point due to formation of
so-called extended van-Hove singularities with increasing of the $t^{\prime
} $. We show one of the AFM-dSC solution on the Fig.\ref{GTAU} with the next
nearest-neighbor hopping for the 10\% doping level and $\beta =50$ eV$^{-1}$%
. The resulting local magnetic moment is $M$=0.28 $\mu _{B}$ and the d-SC order
parameter $F(0)$=0.036. One can see that the superconduction order parameter
is really of the d$_{x^{2}-y^{2}}$ symmetry since diagonal elements ($F^{00}$%
) as well as the next nearest-neighbors elements ($F^{02}$) are all equal to
zero and only the nearest neighbor superconducting Green functions ($F^{01}$%
) are non-zero and change the sign for $F_{x}$ and $F_{y}$ components. The
normal local Green function ($G^{00}$) (plotted for the spin-up atom in the
Fig.\ref{GTAU}) as well as ($G^{02}$) are spin-split, while the
nearest-neighbor Green function ($G^{01}$) has no spin-splitting due to AFM
spin symmetry (see Fig.\ref{AFMSC}). The absence of magnetic polarization in
the non-diagonal G-function along the x(y)-directions suppress the magnetic
pair-breaking and makes the AFM-dSC coexistence possible.

%------------ end of article ------------------->>

%% optional
\section{Summary}

The spectral density functional (SDF) approach allows us to study  correlation
effects in solids based on realistic electronic structure calculations.
Among all possible applications we have chosen the magnetism of transition
metals and high-temperature superconductivity. From these two examples one
can see already all the main advantages of the new approach in comparison
with standard density functional theory. First, we can describe now the
spectral density transfer phenomena (e.g., the formation of 6 eV satellite
in Ni), the quasiparticle damping and other effects connected with the
frequency dependence of the self-energy; they are absent completely not only
in the DF approach but also in the Hartree-Fock, LDA+U, or self-interaction
corrections approximation (see, e.g., \cite{revU}). Second, we can describe
adequately the contribution of the Bose degrees of freedom (e.g., spin
fluctuations) to the electronic structure and thermodynamic properties. In
the DF-based calculations the temperature is really taken into account only
via the thermal expansion and the Fermi distribution function \cite{Jarlborg}%
. It was the main reason for the failure of the standard band theory for the
description of finite-temperature effects in magnetic metals. We show that
the SDF gives a satisfactory solution of this problem.

Most of real applications of the SDF approach are connected
with the single-site dynamical mean field theory. At the same
time, for a number of problems this can be insufficient and
some generalizations to take into account the non-local effects
are necessary, for example, cluster ones. The first attempts of
such generalizations are  already  leading to some interesting
results in the theory of high-T$_{c}$ superconductivity \cite
{jarrell,LK} but a  lot of additional work is required in the
area of cluster dynamical mean field theories  to reach the level
of understanding that was reached in single site DMFT.

%% optional
\begin{acknowledgments}

The work was supported by the Netherlands Organization for
Scientific Research (NWO project 047-008-16) and partially
supported by Russian Basic Research Foundation (grant
00-15-96544). GK is supported by the NSF under grantNSF DMR
0096462 , by the division of Basic Sciences of the Department of
Energy under grant DE-FG02-99ER45761, and by the ONR under \ grant
N000149910653.
\end{acknowledgments}

%% appendix optional
\appendix{Multi-orbital QMC scheme}

The multi-orbital DMFT problem and general cluster DMFT scheme\ can be
reduced to the general impurity action:

\[
S=-\int_{0}^{\beta }d\tau \int_{0}^{\beta }d\tau ^{\prime
}\sum_{i,j}c_{i}^{+}(\tau ){\cal G}_{ij}(\tau -\tau ^{\prime })c_{j}(\tau
^{\prime })+\frac{1}{2}\int_{0}^{\beta }d\tau \sum_{i,j}n_{i}(\tau
)U_{ij}n_{j}(\tau )
\]

where $i=\{m,\sigma \}$ - orbital (site) and spin. Without spin-orbital
coupling we have: ${\cal G}_{ij}={\cal G}_{m,m^{\prime }}^{\sigma }\delta
_{\sigma \sigma ^{\prime }}.$

The auxiliary fields Green-function QMC use the discrete
Hubbard-Stratanovich transformation inroduced by Hirsch\cite{HirschHS}

\[
\exp \{\Delta \tau U_{ij}[n_{i}n_{j}-\frac{1}{2}(n_{i}+n_{j})]\}=\frac{1}{2}%
\sum_{S_{ij}=\pm 1}\exp \{\lambda _{ij}S_{ij}(n_{i}-n_{j})\}
\]
where $S_{ij}(\tau )$ are the auxiliary Ising fields for each pair of
orbitals and time slice with the strength:

\[
\lambda _{ij}={\rm arccosh}[\exp (\frac{\Delta \tau }{2}U_{ij})]
\]

Using Hirsch transofrmation one can integrated out fermionic fields in the
path integral\cite{GKKR} and resulting partition function and Green function
matrix have the following form:

\begin{eqnarray*}
Z &=&\frac{1}{2^{N_{f}L}}\sum_{S_{ij}(\tau )}\det [\widehat{G}^{-1}(S_{ij})]
\\
\widehat{G} &=&\frac{1}{Z}\frac{1}{2^{N_{f}L}}\sum_{S_{ij}(\tau )}\widehat{G}%
(S_{ij})\det [\widehat{G}^{-1}(S_{ij})]
\end{eqnarray*}
where $N_{f}$ is the number of Ising fields, $L$ is the number of time slices, and
 $\widehat{G}(S_{ij})$ is the Green function in the auxiliary Ising
fields:

\begin{eqnarray*}
G_{ij}^{-1}(S) &=&{\cal G}_{ij}^{-1}+\Delta _{i}\delta _{ij}\delta _{\tau
\tau ^{\prime }} \\
\Delta _{i} &=&(e^{V_{i}}-1) \\
V_{i}(\tau ) &=&\sum_{j(\neq i)}\lambda _{ij}S_{ij}(\tau )\sigma _{ij}
\end{eqnarray*}
here we introduce the generalized Pauli matrix:

\[
\sigma _{ij}=\{
\begin{array}{c}
+1,i<j \\
-1,i>j
\end{array}
\]

For efficient calculation of the Green function in arbitrary configuration
of Ising fields $G_{ij}(S)$ we use the following Dyson equation \cite{Hirsch}%
:

\[
G^{\prime }=[1+(1-G)(e^{V^{\prime }-V}-1)]^{-1}G
\]

The QMC important sampling scheme allowed us to integrate over the Ising
fields with the abs($\det [\widehat{G}^{-1}(S_{ij})]$ ) \ as a stochastic
weight\cite{Hirsch,GKKR}. For a single spin-flip $S_{ij},$ the determinant
ratio is calculated as following: \
\begin{eqnarray*}
\det [\widehat{G}]/\det [\widehat{G^{\prime }}]%
&=&R_{i}R_{j}-R_{ij} \\
R_{i} &=&1+[1-G_{ii}(\tau ,\tau )]\Delta _{i}(\tau ) \\
R_{j} &=&1+[1-G_{jj}(\tau ,\tau )]\Delta _{j}(\tau ) \\
R_{ij} &=&G_{ij}(\tau ,\tau )\Delta _{j}(\tau )G_{ji}(\tau ,\tau )\Delta
_{i}(\tau )
\end{eqnarray*}

and the Green function matrix updated in the standard maner\cite{Hirsch,GKKR}%
:

\begin{eqnarray*}
G_{i_{1}j_{2}}^{\prime }(\tau _{1},\tau _{2}) &=&G_{i_{1}j_{2}}(\tau
_{1},\tau _{2})+[G_{i_{1}i}(\tau _{1},\tau )-\delta _{i_{1}i}\delta _{\tau
_{1},\tau }]\Delta _{i}(\tau )/R_{_{i}}(\tau )G_{ij_{2}}(\tau ,\tau _{2}) \\
G_{i_{1}j_{2}}^{new}(\tau _{1},\tau _{2}) &=&G_{i_{1}j_{2}}^{\prime }(\tau
_{1},\tau _{2})+[G_{i_{1}j}^{\prime }(\tau _{1},\tau )-\delta
_{i_{1}j}\delta _{\tau _{1},\tau }]\Delta _{j}(\tau )/R_{_{j}}(\tau
)G_{jj_{2}}^{\prime }(\tau ,\tau _{2})
\end{eqnarray*}

%
% Bibliography made with BibTeX:
%\begin{chapthebibliography}{<widest bib entry>}
%\bibitem[optional]{symbolic name}

\end{document}